\documentclass[british,amssymb,aps,prd,twocolumn,amsmath,floats,floatfix,superscriptaddress,nofootinbib,%
               showkeys,showpacs,groupedaddress]{revtex4-1}
\usepackage{amssymb}
\usepackage{amsmath}
\usepackage{verbatim}
\usepackage{mathrsfs}
\usepackage{amsfonts}
\usepackage{latexsym}
\usepackage{epsfig}
\usepackage{color}
\usepackage[usenames,dvipsnames]{xcolor}
\usepackage{graphicx,subfigure}
\usepackage{units}
\usepackage{envmath}
\usepackage{natbib}
\usepackage{ctable}
\usepackage{soul}
\usepackage[unicode=true,pdfusetitle,
 bookmarks=true,bookmarksnumbered=false,bookmarksopen=false,
 breaklinks=false,pdfborder={0 0 1},backref=false,colorlinks=false]{hyperref}
\usepackage{mathtools}
\begin{document}
\definecolor{orange}{rgb}{0.9,0.45,0}
\def\CovDev{D}
\def\Res{{\mathcal R}}
\def\Gammaflat{\hat \Gamma}
\def\metricflat{\hat \gamma}
\def\Dflat{\hat {\mathcal D}}
\def\part_n{\partial_\perp}
\def\Lie{\mathcal{L}}
\def\A{\mathcal{X}}
\def\Aphi{\A_{\phi}}
\def\hAphi{\hat{\A}_{\phi}}
\def\E{\mathcal{E}}
\def\Ham{\mathcal{H}}
\def\M{\mathcal{M}}
\def\R{\mathcal{R}}
\def\p{\partial}
\def\hg{\hat{\gamma}}
\def\hA{\hat{A}}
\def\hD{\hat{D}}
\def\hE{\hat{E}}
\def\hR{\hat{R}}
\def\hcA{\hat{\mathcal{A}}}
\def\hDelt{\hat{\triangle}}
\def\na{\nabla}
\def\dif{{\rm{d}}}
\def\non{\nonumber}
\newcommand{\erf}{\textrm{erf}}
\renewcommand{\t}{\times}
\long\def\symbolfootnote[#1]#2{\begingroup%
\def\thefootnote{\fnsymbol{footnote}}\footnote[#1]{#2}\endgroup}
\title{Cosmological Inflation in $f(R,T)$ Gravity with Chern-Simons Correction}
\author{Maryam Shiravand$^{1,2}$}
\email{ma\_shiravand@kntu.ac.ir}
\author{Saeed Fakhry$^{3}$}
\email{saeed.fakhry@uv.es}
\author{Mehrdad Farhoudi$^{1}$}
\email{m-farhoudi@sbu.ac.ir}
 \affiliation{${}^{1}$Department of Physics,
              Shahid Beheshti University, 1983969411, Tehran, Iran}
\affiliation{${}^{2}$Department of Physics, K.N. Toosi University
             of Technology, P.O. Box 15875-4416, Tehran, Iran}
\affiliation{${}^{3}$Department of Astronomy and
             Astrophysics, University of Val\`{e}ncia, 19.46100 Burjassot, Val\`{e}ncia, Spain}

\date{July 20, 2026}
\begin{abstract}
\noindent
 We investigate cosmological inflation within the framework of
a linear form of $f(R, T)$ gravity that incorporates an inflaton
scalar field augmented by a Chern-Simons correction induced by
aspects of quantum gravity. Utilizing the FLRW metric, we derive
the modified Friedmann equations under the slow-roll
approximations. We consider two specific forms of the Chern-Simons
coupling function, trigonometric and exponential, each paired with
the choice of an inflaton potential. Then, we define the essential
slow-roll parameters and acquire their required expressions in the
proposed model. Subsequently, we compute the scalar spectral
index, the tensor spectral index, and the tensor-to-scalar ratio.
By adequately constraining the free parameters, the proposed model
provides accurate predictions for these inflationary observables
that are in good agreement with the Planck $2018$ data.
Furthermore, the model predictions for the Chern-Simons
exponential coupling function impose a stronger limit on the value
of the tensor-to-scalar ratio and also provide good agreement with
the joint Planck, BK$15$ and BAO data. Meanwhile, for comparative
analysis and better comparison of the model motivations, we also
examine the model without the Chern-Simons correction, without the
linear form of $f(R,T)$ gravity, and with a non-linear form of
$f(R,T)$ gravity. As a general conclusion, the obtained findings
indicate that the inclusion of the Chern-Simons correction
approximately refines the values of the tensor spectral index and
tensor-to-scalar ratio in the context of the linear form of
$f(R,T)$ gravity.
\end{abstract}

\maketitle
\vspace{8cm}

\section{Introduction}
In recent decades, a plethora of theoretical and observational
studies have been conducted to elucidate the dynamics of the
Universe. These studies largely corroborate the standard
cosmological model, known as the $\Lambda$CDM model, which is
based on general relativity (GR)~\cite{Ferreira:2019xrr}.
Observations of the cosmic microwave background (CMB) radiation,
facilitated by various surveys, provide important insights into
the formation and evolution of the Universe, thereby bolstering
the $\Lambda$CDM model~\cite{WMAP:2003elm,WMAP:2006bqn,
WMAP:2008lyn, Hinshaw2013,Planck:2018jri,Planck:2018vyg}. However,
there are three profound enigmas in contemporary cosmology, namely
the nature of dark matter, the dark energy conundrum in the
context of the late Universe, and the phenomena occurring during
the early Universe, all of which are still unclear. Indeed,
cosmology is presently at a juncture where the availability of
extensive observational data necessitates robust theoretical
frameworks for explanation.

On the other hand, some challenging concepts, such as the
flatness, horizon, and monopole problems that arise in the big
bang cosmology, remain unresolved in the $\Lambda$CDM model, see,
e.g., Ref.~\cite{Coley:2019yov}. To address these issues, the
concept of cosmological inflation has been proposed, see, e.g.,
Refs.~\cite{Starobinsky1980, guth1981, lidd1982,
Albrecht1982,Liddle1999, Baumann2009, Martin:2013tda,
Odintsov2023,Kallosh}. Indeed, cosmic inflation is a hypothetical
yet well-supported phase of rapid accelerated expansion in the
very early Universe. The inflationary framework is usually
explained by introducing a scalar field called the inflaton, which
is governed by a specific potential. The quantum fluctuations of
the inflaton can instigate an inflationary era and account for the
origin of large-scale structures in the Universe.

Cosmological inflation generates density perturbations whose
effects are observable in the measurements of the CMB temperature
anisotropies~\cite{2dFGRS:2001csf, WMAP:2003syu, SDSS:2003tbn,
SDSS:2003eyi}. In fact, inflation is likely responsible for
creating the primordial inhomogeneities that serve as seeds for
the large-scale structure of the Universe, i.e., the
near-scale-invariant spectrum of primordial density perturbations
recently observed by the Planck satellite at a high level of
precision~\cite{Planck:2018jri}. To sustain a sufficiently
prolonged inflationary era, the slow-roll conditions must be
imposed on the inflaton field, under which the kinetic energy of
the inflaton field must be negligible compared to its potential
energy. Hence, under such a common assumption, the inflaton scalar
field should evolve very slowly. The various potentials that can
describe an inflationary era have been extensively studied and
constrained through the CMB anisotropy
measurements~\cite{Hossain:2014coa, Martin:2013nzq, Geng:2015fla,
Martin:2015dha, Huang:2015cke}. Although GR has provided the most
accurate predictions to describe many cosmological
phenomena~\cite{Will:2005va}, it fails to adequately account for
the dark sectors of the Universe whose influence on the dynamics
of the Universe is compatible with observational
data~\cite{Ishak2019}. The limitations of GR have led to the
exploration of alternative theories of
gravity~\cite{Jackiw:2003pm,Farhoudi2006,Alexander:2009tp, reff19,
Felice, Sotiriou2010,capozziello2011,Harko:2011kv,Bueno2016,
khosravi2016,quiros2019} that mostly aim to overcome the
shortcomings of GR~\cite{Atazadeh2008, Yunes:2009hc,clifton,
Farajollahi2012, Bahrehbakhsh-2013, Shabani2014, Joyce2015,zare1,
Haghani:2017yjk,Nojiri:2017ncd, Haghani:2015zga, Mishra2020,
Y:2021ybx, Odintsov:2022hxu,Nashed:2023mqh,Fakhry:2023fbt,
Daniel:2024lev,Fakhry:2024fwz,arXiv240811995F}. In this regard,
various modified theories of gravity have been considered within
the context of inflationary cosmology, with the aspiration that
their predictions may be in better agreement with observational
data~\cite{Satoh:2010ep,Alexander:2011hz,
Garrison:2012uxi,Myrzakulov:2015qaa,DeLaurentis:2015fea,
Kanti:2015pda, Sebastiani:2016ras, Tirandari:2017nzy,saba,RSFM,
Chakraborty:2018scm,Bernardo,
Kausar2019,Granda2019,Bhattacharjee:2020jsf,Odintsov:2020sqy,
Jalalzadel-et-al, Gamonal:2020itt,
TQ.Do,baffou,Venikoudis:2021oee, faraji,Pozdeeva, Bhattacharjee,
Zhang:2021ppy, Shiravand:2022ccb, Chen:2022dyq,Bagherian:2022mau,
Shiravand:2024ayw,Shiravand2025,Ahghari2026}.

Currently, due to observational limitations, the arena of
inflation is in a phase of hypothesizing and fitting models that
may coherently explain cosmological phenomena. In the context of
early cosmological epochs, the approach transitions from the
domain of classical gravity to the yet-to-be-explored era of
quantum gravity, which is hypothesized to govern small-scale
physics and unify all fundamental forces in
nature~\cite{Gong:2010gs, Secrest:2022uvx, Cao:2023jbo}. Between
the classical and quantum gravity regimes, a period of rapid
acceleration with a pseudo-exponential expansion rate is believed
to have occurred, known as the inflationary
period~\cite{Agullo:2012sh}. However, the effective Lagrangian of
inflation remains undetermined by the existing data. Thus, while
the inflationary epoch can be considered a classical phase of the
Universe described by a four-dimensional spacetime, it is
plausible that the quantum epoch directly influences the effective
inflationary Lagrangian~\cite{German:2020dih, German:2020roi}. In
this respect, some of the effective corrections to the
inflationary Lagrangian may involve higher-curvature terms, such
as those found in $f(Q, T)$
gravity~\cite{Shiravand:2022ccb,Shiravand:2024ayw}, $f(R, T)$
gravity~\cite{Bhattacharjee:2020jsf, Gamonal:2020itt, baffou,
Siggia}, the Gauss-Bonnet correction~\cite{Y:2021ybx,
Daniel:2024lev, Satoh:2010ep, Kanti:2015pda,
Chakraborty:2018scm,Pozdeeva,Odintsov:2020sqy,
Venikoudis:2021oee,Ahghari2026,slowroll7,jiang8, reff27, reff30,
reff37,Zhu2025,Yogesh}, and the Chern-Simons
corrections~\cite{Jackiw:2003pm,Alexander:2009tp,
Yunes:2009hc,Haghani:2017yjk,Odintsov:2022hxu, Nashed:2023mqh,
Daniel:2024lev,Satoh:2010ep, Alexander:2011hz, Garrison:2012uxi,
Venikoudis:2021oee,Bagherian:2022mau}.

Among the modified theories of gravitation, the $f(R,T)$ gravity
theory integrates dependencies on both the Ricci scalar, $R$, and
the trace of the energy-momentum tensor, $T$. It provides a
straightforward extension of GR and establishes a direct
interaction between matter and curvature, a feature not~present in
the standard GR and $f(R)$ gravity, see, e.g.,
Refs.~\cite{Artymowski:2014gea,Sharma:2022tce}. Such an
enhancement provides a more comprehensive framework for describing
gravitational phenomena by incorporating the contribution of
matter directly into the gravitational action. A key advantage of
$f(R, T)$ gravity in cosmology is its ability to address both dark
energy and dark matter problems, see, e.g.,
Refs.~\cite{Chakraborty:2012kj, Zaregonbadi:2016xna,
Errahmani:2024ran}. By introducing additional degrees of freedom,
$f(R, T)$ gravity can provide a mechanism for deriving cosmic
acceleration without the need for a cosmological constant and
potentially explain the observed late-time acceleration of the
Universe. Furthermore, this theory enables the exploration of
interactions between matter and curvature, yielding novel insights
into the evolution of large-scale structures and the dynamics of
the early Universe~\citep{phsp,Rudra:2020nxk, Nagpal:2021akx,
Bhardwaj:2021ymy, Goncalves:2022ggq}. The flexibility in the
functional form of $f(R, T)$ allows the construction of models
that can be precisely aligned with observational data, thereby
enhancing the predictive power and explanatory scope of
cosmological models~\cite{Bouali:2023fid}. Consequently, $f(R, T)$
gravity is a promising way to address some of the most critical
questions in contemporary cosmology.

On the other hand, Chern-Simons corrections introduce higher-order
topological terms into the gravitational action. Such extension
introduces additional coupling between matter and curvature. These
terms can potentially cause significant modifications in the
behavior of spacetime in gravitational fields, see, e.g.,
Refs.~\cite{Jackiw:2003pm, Sun:2008uf, Zanelli:2012px,
McNees:2015srl, Haghani:2017yjk, Nakamura:2018yaw,
Odintsov:2022hxu, Daniel:2024lev}. Such adjustments provide
particular utility in cosmology by furnishing a means to address
parity violations and anomalies in gravitational interactions,
which are~not accounted for in the conventional cosmological
framework. Through the incorporation of Chern-Simons corrections,
modified gravity theories can furnish deeper insights into the
dynamics of the Universe, see, e.g., Refs.~\cite{Yunes:2009hc,
Alexander:2009tp, Satoh:2010ep, Venikoudis:2021oee,
Nashed:2023mqh}. Moreover, Chern-Simons corrections contribute to
the development of a more inclusive theory of quantum gravity,
ensuring adherence to fundamental symmetry
principles~\citep{Bonezzi:2014nua, Alexander:2022vpn,
Castro:2023bvo}. Incorporating such corrections augments
theoretical adaptability and allows the formulation of models that
better conform with observational data, thereby enhancing the
explanatory and predictive capabilities of cosmological theories.

Thereby, in this study, we investigate cosmological inflation in
the framework of $f(R,T)$ gravity, while also incorporating
higher-order Chern-Simons terms into the gravitational action,
motivated by aspects of quantum gravity. This extension introduces
an additional coupling between curvature and matter, potentially
leading to intriguing phenomenological consequences for the early
Universe. Hence, in the next section, we model an inflationary
scenario in a linear form of $f(R,T)$ gravity with an inflaton
scalar field in the presence of a Chern-Simons correction. The
choice of linear form of $f(R,T)$ gravity is due to its simplicity
and ability to introduce such a minimal coupling, which allows for
a tractable analytical treatment under the slow-roll
approximations. In fact, it allows to derive explicit expressions
for the slow-roll parameters and inflationary observables, while
still capturing the essential features of $f(R,T)$ gravity. Then,
while considering the slow-roll approximations, we also delve in
the corresponding Friedmann equations for the proposed model. In
Sec.~\ref{sec:iii}, in this analysis, we consider two distinct
forms of the inflaton potential with two variants of the
Chern-Simons coupling function. Then, we compute the key
inflationary observables and constrain the range of the parameters
by systematically comparing the obtained results with
observational data. However, for comparative analysis and better
comparison of results, we check some of the main motivations of
the presented model. Mainly, in this section, we investigate the
model without the Chern-Simons correction and the model without
the linear form of $f(R,T)$ gravity (i.e., by eliminating the
trace of the energy-momentum tensor) for both potentials. Also, in
Sec.~IV, we examine the model with a non-linear form of $f(R,T)$
gravity for the power-law potential and the Chern-Simons
trigonometric coupling function. Finally, in Sec.~V, we scrutinize
the obtained results and summarize the findings.

\section{Inflation in $f(R,T)$ Gravity with Chern-Simons and Inflaton}\label{sec:ii}
Beyond the standard models of inflation including scalar fields,
the inflationary scenario has also been investigated through
alternative theories of gravity. In this section, we examine
slow-roll inflation in the framework of a linear form of $f(R,T)$
gravity while incorporating a Chern-Simons correction and an
inflaton scalar field. Such a combination allows for a more
comprehensive approach to addressing gravitational interactions.
Thus, for the proposed model, we consider the gravitational action
\begin{equation}\label{action}
S\!=\!\int \sqrt{-g}\, d^{4}x \bigg[\dfrac{f(R,
T)}{2\kappa^2}-\dfrac{1}{2}\partial^{a}\phi\partial_{a}\phi-V(\phi)
+L_{[\rm CS]}^{}\bigg],
\end{equation}
where $L_{[\rm CS]}^{}=\nu(\phi)R\widetilde{R}/8$ is the
Chern-Simons correction or the string axionic correction
term~\cite{Hwang2005}. Here, $g$ denotes the determinant of the
metric, the lowercase Latin indices run from zero to three, and
the $\kappa^2$ parameter is $8\pi$ in the Planck units with
$G=1=c$ and $\hbar=1$. Also, $\phi$ is the inflaton scalar field
with the potential $V(\phi)$, and function $\nu(\phi)$ serves as
the Chern-Simons coupling function. The Chern-Simons theorem
indicates parity violation in gravity represented by the
expression $R\widetilde{R} = \epsilon^{abcd}R_{ab}{}^{ef}
R_{cdef}$, which involves the completely antisymmetric Levi-Civita
tensor density, $\epsilon^{abcd}$, in four dimensions and the
Riemann tensor, $R_{abcd}$.

In this study, we choose the linear form $f(R,T) = R + \beta
\kappa^2 T$, where $\beta$ is a dimensionless coupling constant.
We then utilize the spatially flat homogeneous and isotropic
Friedmann-Lema\^{\i}tre-Robertson-Walker (FLRW) metric
\begin{equation}
ds^2=-dt^2+a^2(t)\Big(dx^2+dy^2+dz^2\Big),
\end{equation}
where $a(t)$ is the scale factor. In this vein, we also assume
that the inflaton scalar field inherits the symmetry of spacetime
and depends only on the cosmic time, i.e. $\phi=\phi (t)$.

Now, by varying the action with respect to the metric and
thereafter, using the Hubble parameter, $H = \dot{a}/a$, and the
Ricci scalar in this metric, $R=12H^2+6\dot{H}$, we obtain the
modified Friedmann and Raychadhuri equations, respectively, as
\begin{eqnarray}
\label{h2}3H^2=\dfrac{\kappa^2}{2}\Big[(1+\beta)
\dot{\phi}^2+2(1+2\beta)V\Big]\equiv\kappa^2\rho^{[\rm eff]}
\end{eqnarray}
and
\begin{align}
\label{frw2}3H^2+2\dot{H} =-\dfrac{\kappa^2}{2} \Big[(1+\beta)
\dot{\phi}^2-2(1+2\beta)V\Big]\!\equiv -\kappa^2 p^{[\rm eff]},
\end{align}
where the dot represents the derivative with respect to the cosmic
time, $t$, and $p^{[\rm eff]}$ and $\rho^{[\rm eff]}$ are the
effective pressure and energy densities, respectively. Then,
substituting Eq.~\eqref{h2} into Eq.~\eqref{frw2} yields the
evolution equation for $H$, which provides a complete description
of how the Hubble parameter evolves with time under the influence
of the inflaton scalar field and the model parameter, as
\begin{equation}\label{hdot}
\dot{H}=-\dfrac{\kappa^2}{2}(1+\beta) \dot{\phi}^2.
\end{equation}
Moreover, utilizing Eqs.~\eqref{h2} and \eqref{frw2}, the
characterizing quantities in the effective equation of state
parameter are
\begin{equation}\label{eos}
w^{[\rm eff]}\!=\dfrac{p^{[\rm eff]}}{\rho^{[\rm
eff]}}=-1-\dfrac{2}{3}\dfrac{\dot{H}}{H^2}
=\dfrac{(1+\beta)\dot{\phi}^2-2(1+2\beta)V}{(1+\beta)\dot{\phi}^2+2(1+2\beta)V},
\end{equation}
where, in general, $\beta\neq -1$ and $\neq -1/2$.

In addition, by differentiating Eq.~\eqref{h2} with respect to
time and then substituting Eq.~\eqref{hdot} into it, we can derive
the modified Klein-Gordon equation as\footnote{Obviously, this
equation can be obtained by the variation of the action with
respect to the inflaton scalar field and then using the FLRW
metric.}
\begin{equation}\label{KG}
\ddot{\phi}(1+\beta)+3H(1+\beta)\dot{\phi}+(1+2\beta)V^\prime=0,
\end{equation}
where the prime indicates the derivative with respect to the
inflaton scalar field, $\phi$. It should be noted that although
the Chern-Simons correction contributes the term
\begin{equation}\label{EMofCS}
T^{ab}_{[\rm
CS]}{}^{}=\epsilon^{acde}\left(\nu_{;ef}R^{fb}{}_{cd}-2\,\nu_{,e}R^b{}_{c;d}\right)
\end{equation}
to the gravitational field equations, it has~no direct impact on
the background Friedmann equations due to the FLRW metric.
However, as we will see, it can still affect the power spectrum of
tensor perturbations, indicating the characteristic of the
Chern-Simons correction.

To delve into the inflationary epoch of the Universe, it becomes
imperative to tackle the system of motion equations analytically,
and this endeavor presents considerable challenges. Accordingly,
it is common to utilize the usual slow-roll
approximations/conditions during the inflationary phase. That is,
by employing $\dot{\phi}^2\ll V(\phi)$ and
$\mid\!\ddot{\phi}\!\mid\ll H\mid\!\dot{\phi}\!\mid$,
Eqs.\,\eqref{h2} and \eqref{KG} can be simplified as
\begin{equation}
\label{h2app}3H^2\approx \kappa^2(1+2\beta)V,
\end{equation}
\begin{equation}
\label{phidot}\dot{\phi}\approx
-\dfrac{1+2\beta}{3H(1+\beta)}V^\prime.
\end{equation}
Furthermore, by applying the usual slow-roll approximations, the
effective equation of state parameter \eqref{eos} is minimized to
approximately $w^{[\rm eff]}\approx -1$.

It is worth noting that we have employed the usual slow-roll
approximations for this $f(R,T)$ gravity with Chern-Simons
correction, as in the Einstein-Hilbert action. However, in
sufficiently rich structure of certain modified gravity models,
the validity of the slow-roll approximations warrants careful
scrutiny. Indeed, additional gravitational degrees of freedom or
higher-curvature corrections can, in principle, alter the
background inflationary dynamics, thereby potentially affecting
the reliability of the slow-roll treatment. Hence, it may be
necessary to define additional and appropriate slow-roll
approximations (see, relations~\eqref{NewConditions} and e.g.,
Refs.~\cite{Shiravand:2024ayw,Ahghari2026,Sarkar}). In any case,
these approximations must be such as to ensure that the slow-roll
conditions are consistently met.

However, the current study is considerably simplified by the
specific structure of the adopted theory. Specifically, although
the action includes the Chern-Simons higher-curvature correction,
this term does~not contribute to the background Friedmann
equations. Furthermore, our choice of the linear model,
$f(R,T)=R+\beta \kappa^2 T$, represents the closet extension of GR
within the $f(R,T)$ framework. In this construction, the pure
gravitational sector remains identical to GR. Thereby, the model
maintains the structural simplicity of the background dynamics.
Nevertheless, in this work, the resulting slow-roll dynamics
essentially encompass the leading-order contribution of the
proposed modified gravity framework, and we show that these
conditions are sufficient for inflation to proceed.

Meanwhile, it should be noted that in the Chern-Simons gravity the
emergence of ghost modes is an important consideration, as the
introduction of higher-order terms can lead to instability, see,
e.g., Ref.~\cite{Dyda:2012rj}. However, in this context, in
Ref.~\cite{Bartolo}, they work within an energy range in which
ghost fields are absent. In the context of our model, since the
Chern-Simons term modifies the tensor perturbations but does~not
directly affect the background Friedmann equations or the scalar
perturbations, the risk of ghost modes is reduced. Also, while
ghost modes can arise in certain contexts, the specific coupling
functions and the slow-roll conditions in our model help to
mitigate such risk. However, a full analysis of ghost modes
requires a more detailed study of the perturbative stability of
the theory, which can be considered in future studies,
particularly in the context of an ultraviolet-complete quantum
gravity framework.

On the other hand, in general, the dynamics of inflation can be
encapsulated by a set of six parameters, each of which plays a
crucial role in shaping the evolution of the Universe during the
inflationary epoch. These parameters, known as the slow-roll
parameters, are precisely delineated to capture the intricate
nuances of inflationary behavior and are generally defined
as~\cite{Nojiri:2017ncd,Venikoudis:2021oee,Hwang2005}
\begin{eqnarray}\label{defsr}
 \epsilon_1\! &\equiv & -\dfrac{\dot{H}}{H^2}, \quad
 \epsilon_2\! \equiv -\dfrac{\mid\!\ddot{\phi}\!\mid}{H\!\mid\!\dot{\phi}\!\mid},\quad
 \epsilon_3\! \equiv \dfrac{\mid\!\dot{F}\!\mid}{2H\!\mid\!F\!\mid},\nonumber\\
 \epsilon_4\! &\equiv & \dfrac{\mid\!\dot{E}\!\mid}{2H\!\mid\!E\!\mid}, \
 \epsilon_5\! \equiv \dfrac{ \mid\!\dot{F}+Q_a\!\mid}{2H\!\mid\!F\!\mid}, \
 \epsilon_6\! \equiv\bigg|\!
 \sum_{i=L,R}\dfrac{Q_t^{\prime}(\lambda_i,\phi)\dot{\phi}}{2H Q_t
 (\lambda_i,\phi)}\!\bigg|,\cr &
\end{eqnarray}
where $F \equiv \kappa^{-2}(\partial f/\partial R)$, $E \equiv
F\left[1+3 \dot{F}^2/(2F\dot{\phi}^2)\right]$,
$Q_a=-8\dot{\xi}(\phi)H^2$ (where $\xi(\phi)$ is the coupling
scalar function to the Gauss-Bonnet term, which we do~not consider
it in the presented action~\eqref{action}, i.e., in the presented
model $Q_a=0$), and the parameter $Q_t$ arising from the
Chern-Simons correction for tensor perturbations is defined as
\begin{equation}\label{Qt}
Q_t (\lambda_i,\phi)\equiv F+2\dfrac{k
\lambda_i\,\nu^\prime\dot{\phi}}{a}.
\end{equation}
Here, $\lambda_i$ denotes the polarization of fundamental
gravitational waves, characterized by the wave number $k$, which
takes the values $\lambda_{\rm L}^{}=-1$ and $\lambda_{\rm
R}^{}=1$ for the left and right polarizations of the gravitational
waves, respectively. However, the relation $k/a=H$ can be used in
the analysis of the inflationary quantities at the first horizon
crossing.  Note that, in the absence of the Chern-Simons
correction, although relation \eqref{Qt} gives a constant value,
the parameter $\epsilon_6$ is zero as expected. Moreover, the
parameters $\epsilon_3$, $\epsilon_4$, and $\epsilon_5$ depend on
the cosmic time derivative of $\partial f(R,T)/\partial R$, which
in this context is considered constant for the linear form of
$f(R,T)$ gravity, i.e., in this case $E=F=\kappa^{-2}$. Hence, in
the proposed model, these parameters are zero and do~not
contribute to the corresponding inflationary dynamics. However,
they are relevant in more general cases where $f(R,T)$ is
non-linear.

It is well-known that inflation occurs and persists for a
considerable duration to address conventional cosmological
challenges when the condition $|\epsilon_i| \ll 1$ is
satisfied~\cite{Martin:2013tda}. In addition, inflation terminates
when the initial slow-roll parameters reach unity at the end of
inflation, in particular when $\epsilon_1|_{\rm end}^{} = 1$. Now,
utilizing Eqs. \eqref{hdot}, \eqref{h2app}, \eqref{phidot} into
the definitions of slow-roll parameters \eqref{defsr}, while using
the slow-roll conditions, we can express their non-zero ones in
terms of the potential and the Chern-Simons coupling scalar
function as
\begin{equation}\label{eps1}
\epsilon_1\approx\dfrac{1}{2\kappa^2(1+\beta)}\left(\dfrac{V^\prime}{V}\right)^2,
\qquad
\epsilon_2\approx\dfrac{1}{\kappa^2(1+\beta)}\left(\dfrac{V^{\prime\prime}}{V}\right),
\end{equation}
\begin{equation}
\epsilon_6\!\approx\!\dfrac{\kappa^2
\left(1+2\beta\right)^2\nu^\prime V^{\prime
2}\left(\nu^{\prime\prime}V^\prime +\nu^\prime
V^{\prime\prime}\right)}{\left(1+\beta\right)V\left\lbrace
9\left(1+\beta\right)^2/4-\big[\kappa^2\left(1
+2\beta\right)\nu^\prime V^\prime\big]^2\right\rbrace}.
\label{eps6}
\end{equation}

Also, the e-folding number is a significant parameter in
quantifying the extent of inflationary expansion. It represents
the natural logarithm of the scale factor and provides a measure
of how much the Universe expanded during the inflationary phase.
This parameter is essential for understanding the dynamics and
duration of inflation and is defined as
\begin{equation}\label{e-folding}
N\equiv \ln \( \dfrac {a_{\rm end}}{a}\) =\int_{t}^{t_{\rm end}}
H{\rm d}t,
\end{equation}
where the subscript `end' indicates the value of quantities at the
end of inflation. Then, by inserting Eqs.~\eqref{h2app} and
\eqref{phidot} into relation~\eqref{e-folding}, it reads
\begin{equation}\label{Neq}
N(\phi)= \int_{\phi}^{\phi_{\rm end}}\dfrac{H}{\dot{\phi}}{\rm
d}\phi \approx -\kappa^2(1+\beta) \int_{\phi}^{\phi_{\rm
end}}\dfrac{V}{V^\prime}{\rm d}\phi.
\end{equation}

\section{Chern-Simons Coupling Function and Inflaton Potential}\label{sec:iii}
Up to here, we have~not specified the Chern-Simons coupling
function and the inflaton potential. In the following subsections,
we consider two distinct forms of the inflaton potential with two
variants of the Chern-Simons coupling function. Then, we analyze
and compute the inflationary observables and compare the results
with observational data. However, for comparative analysis and to
better compare the results obtained, for the both inflaton
potentials, we also examine two main motivations of the presented
model in the following subsections. Namely, we investigate the
model without the Chern-Simons correction and the model without
the linear form of $f(R,T)$ gravity (i.e., by eliminating the
trace of the energy-momentum tensor).

\subsection{Power-Law Potential}
As the first case, we consider one of the simplest models, i.e.
the power-law potential of the form
\begin{equation}\label{V1}
V(\phi)=v \phi^n,
\end{equation}
where the non-zero parameters $n$ and $v$ are constants, resulting
in a chaotic inflation scenario~\cite{Linde:1983gd,
Pavluchenko:2003ft}. Obviously, the dimension $v$ for setting the
dimension $V(\phi)$ depends on the value of $n$.

\subsubsection{\bf The Model with Chern-Simons Trigonometric Coupling Function}

In this case, we also consider the coupling function related to
the Chern-Simons correction in the trigonometric form
\begin{equation}\label{nu1}
\nu(\phi)=\Lambda \sin\(\dfrac{\phi}{\phi_1}\),
\end{equation}
where $\Lambda$ and $\phi_1$ are constant parameters.

In this case, the slow-roll parameters \eqref{eps1} and
\eqref{eps6} become
\begin{eqnarray}
\epsilon_1 &\approx & \dfrac{n^2}{2\kappa^2(1+\beta)\phi^2},
\qquad\quad
\epsilon_2 \approx\dfrac{n(n-1)}{\kappa^2(1+\beta)\phi^2}, \label{eps1p} \\
\epsilon_6 &\approx &
\dfrac{n^3\cos\left(\frac{\phi}{\phi_1}\right)\left[\kappa\Lambda
v
\left(1+2\beta\right) \phi^{n-2}\right]^2}{\phi_1(1+\beta)} \nonumber\\
           &&\hspace*{-1cm} \times \dfrac{(n-1)\phi_1 \cos\left(\frac{\phi}{\phi_1}\right)
           -\phi \sin\left(\frac{\phi}{\phi_1}\right)}{9\phi_1^2 (1+\beta)^2/4-\left[n \kappa^2 \Lambda v
           \left(1+2\beta\right) \phi^{n-1}
           \cos\left(\frac{\phi}{\phi_1}\right)\right]^2 }. \nonumber\\
 \label{eps6p}
\end{eqnarray}
Then, by applying the condition of the end of inflation,
specifically $\epsilon_1(\phi_{\rm end}) = 1$, we obtain the value
of the inflaton scalar field at the end of inflation as
\begin{equation}
\label{phiend}\phi_{\rm
end}^2\approx\dfrac{n^2}{2\kappa^2\(1+\beta\)}.
\end{equation}
\begin{figure*}[t!]
\centering \subfigure[]{
\includegraphics[width=0.31\textwidth]{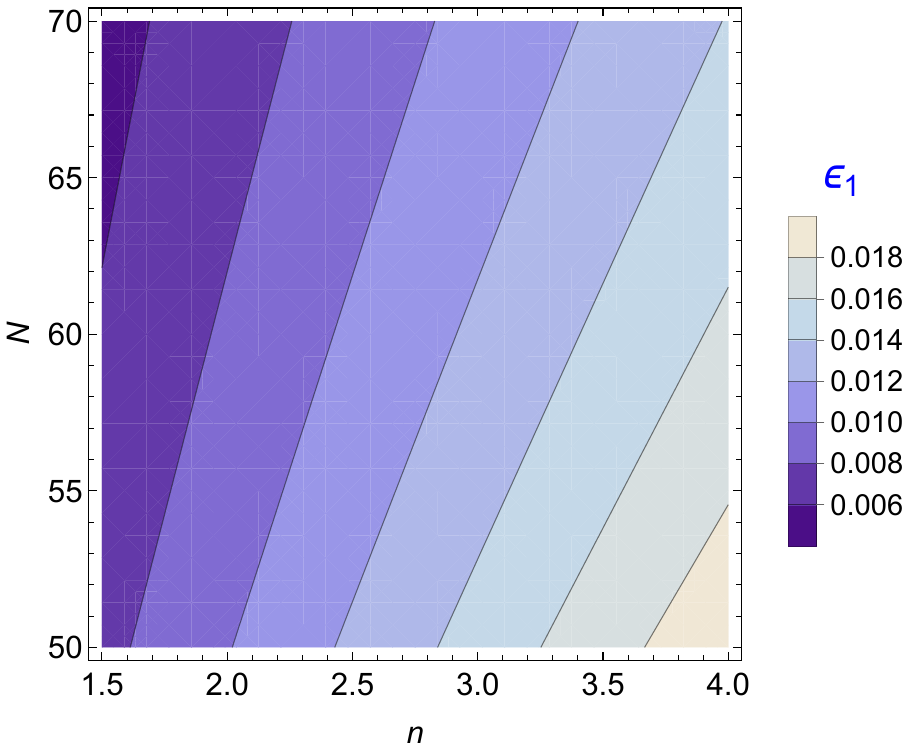}
\label{a1} } \subfigure[] {
\includegraphics[width=0.31\textwidth]{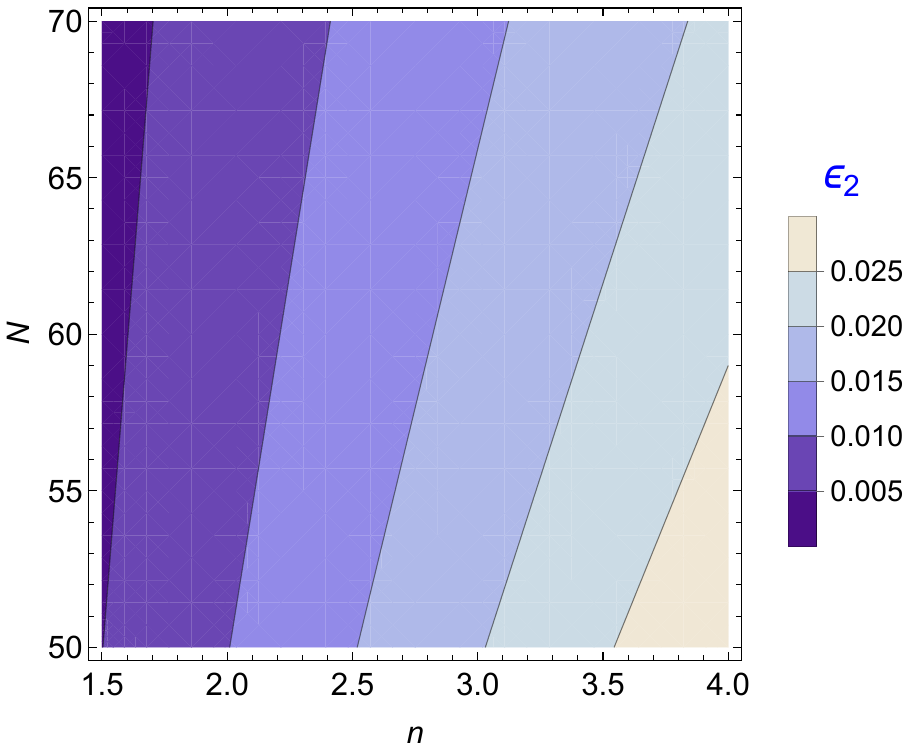}
\label{b1} } \subfigure[] {
\includegraphics[width=0.325\textwidth]{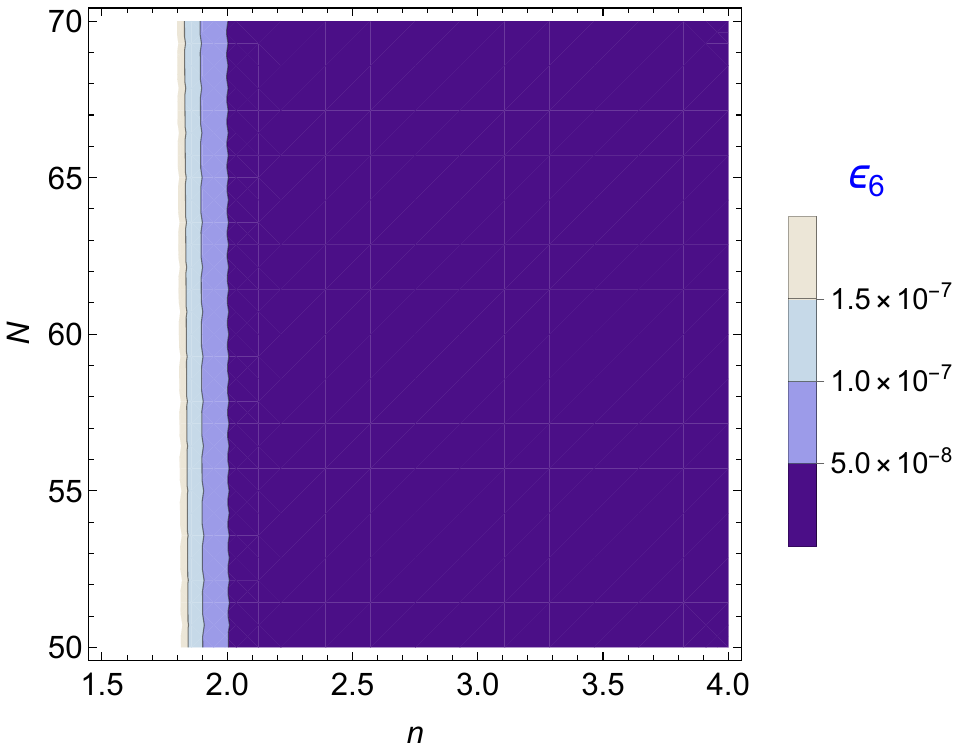}
\label{c1} }
 \caption{\label{fig1} [color online] The slow-roll parameters have been plotted in terms of the e-folding number,
 $N$, and the potential power, $n$, with the values of $\beta=10^{5}$, $\Lambda=0.1$,
 $\phi_1=1/\kappa^2$, $v=1/(8 \pi)^2$, and $\kappa^2=8 \pi$, in the range $n=[1.5,4]$ and
 $N=[50,70]$, where (a) shows the range of parameters $\epsilon_1$, (b) $\epsilon_2$,
 and (c) $\epsilon_6$.}
\end{figure*}
\begin{figure*}[t!]
\centering \subfigure[]{
\includegraphics[width=0.31\textwidth]{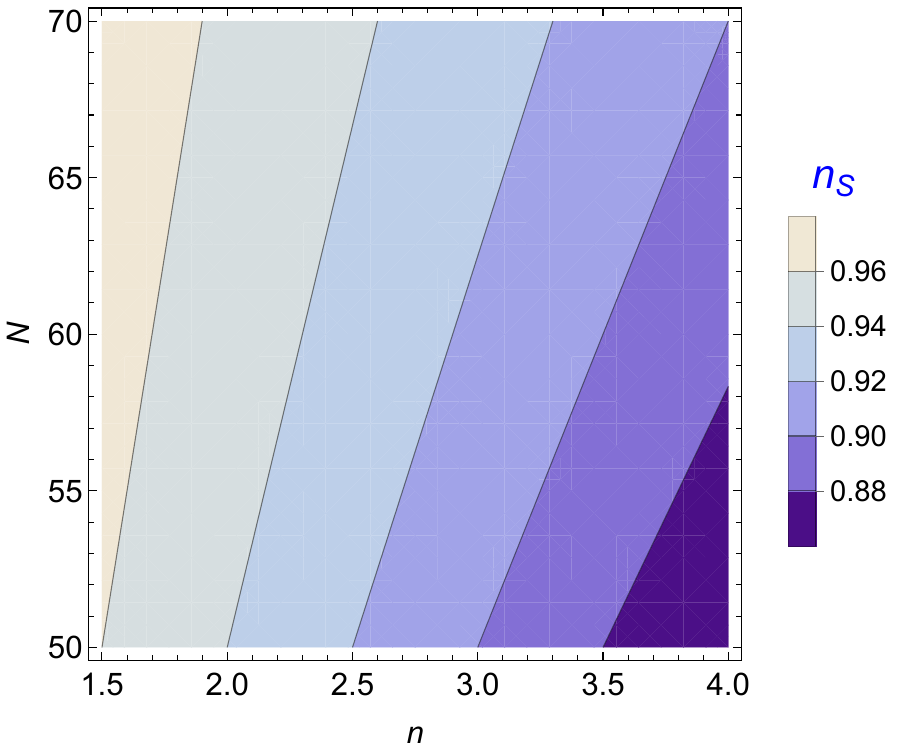}
\label{a2} } \subfigure[] {
\includegraphics[width=0.32\textwidth]{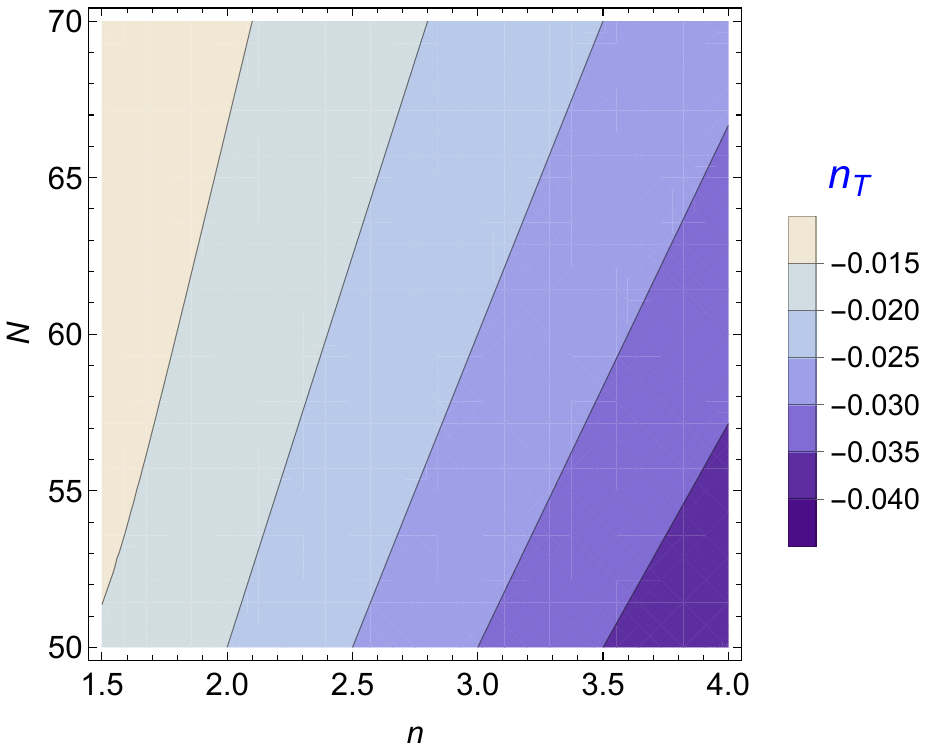}
\label{b2} } \subfigure[] {
\includegraphics[width=0.31\textwidth]{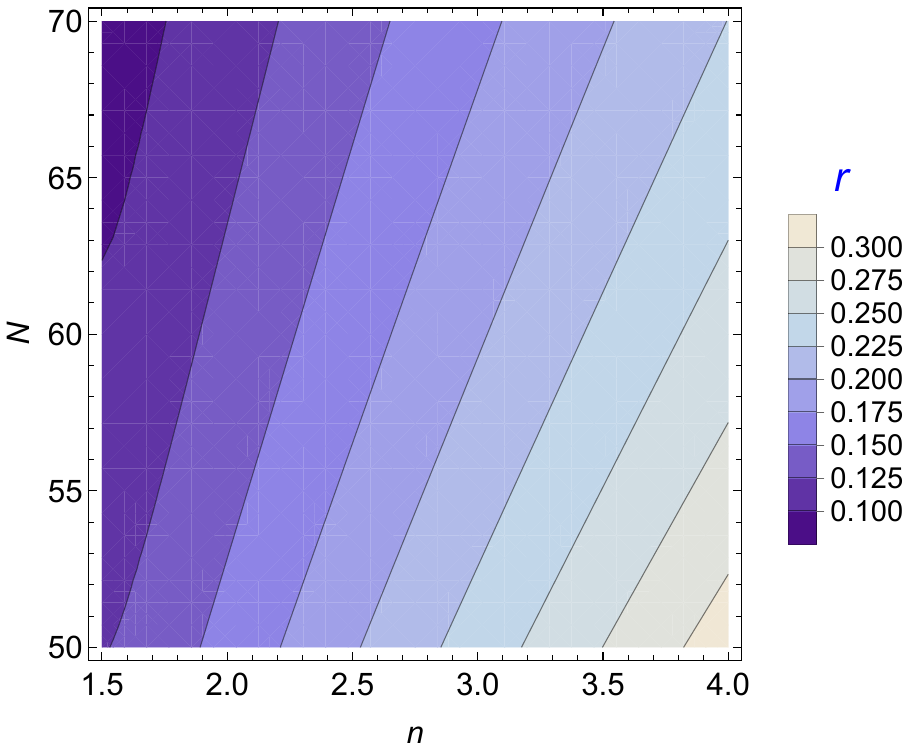}
\label{c2} }
 \caption{\label{fig2} [color online] The inflationary observables have been plotted in terms of the e-folding number,
 $N$, and the potential power, $n$, with the values of $\beta=10^{5}$, $\Lambda=0.1$,
 $\phi_1=1/\kappa^2$, $v=1/(8 \pi)^2$ and $\kappa^2=8 \pi$, in the range $n=[1.5,4]$ and $N=[50,70]$,
 where (a) shows the range of the scalar spectral index, $n_{\rm S}^{}$, (b) the tensor
 spectral index, $n_{\rm T}^{}$, and (c) the tensor-to-scalar ratio, $r$.}
\end{figure*}
Also, by substituting the potential function \eqref{V1} into
relation \eqref{Neq}, we get the e-folding number for this case as
\begin{align}
N(\phi)\approx & -\dfrac{\kappa^2(1+\beta)}{n}\int_{\phi}^{\phi_{\rm end}} \phi {\rm d}\phi,\nonumber\\
&=\frac{\kappa^2(1+\beta)}{2n}\left(\phi^2-\phi_{\rm
end}^2\right).
\end{align}
Subsequently, using relation \eqref{phiend}, we obtain the value
of the inflaton scalar field at the first horizon crossing as
\begin{equation}\label{phiTo2}
\phi^2\approx\dfrac{n\(n+4N\)}{2\kappa^2\(1+\beta\)}.
\end{equation}
Utilizing the above relation, relations~\eqref{eps1p} and
\eqref{eps6p} are obtained as
\begin{eqnarray}
\epsilon_1 &\approx & \dfrac{n}{n+4N}, \qquad\quad
\epsilon_2 \approx\dfrac{2\(n-1\)}{n+4N}, \label{eps11p} \\
\epsilon_6 &\approx & \dfrac{n^3\cos(\sqrt{{\rm
B}})\left[\kappa\Lambda v
\left(1+2\beta\right)\right]^2\left(\phi_1^2{\rm B}\right)^{n-2}}{\phi_1(1+\beta)} \nonumber\\
           && \times \dfrac{(n-1)\phi_1 \cos(\sqrt{{\rm B}})
           -\sqrt{\phi_1^2{\rm B}} \sin(\sqrt{{\rm B}})}{{\rm
           C}}.
 \label{eps66p}
\end{eqnarray}
where ${\rm B}$ and ${\rm C}$ are defined as
\begin{equation}
{\rm B}\equiv \dfrac{n\left(n+4N\right)}{2\kappa^2\phi_1^2
\left(1+\beta\right)},
\end{equation}
\begin{equation}
{\rm C}\!\equiv\! 9\phi_1^2 (1\!+\!\beta)^2\!/4\!-\!\left[n
\kappa^2 \Lambda v \left(1\!+\!2\beta\right)
   \cos(\sqrt{{\rm B}})\right]^2\!\!\left(\phi_1^2{\rm B}\right)^{n-1}\!.
\end{equation}

As is evident from the analysis, while the parameters $\epsilon_1$
and $\epsilon_2$ show dependence only on $n$ and $N$, the
parameter $\epsilon_6$ additionally shows more dependence on the
specific model parameters $\beta$, $v$, $\Lambda$, and $\phi_1$.
This augmented dependence stems from the inclusion of the
Chern-Simons correction in action \eqref{action}, which differs
from the behavior observed in $f(R,T)$ gravity. In
Fig.~\ref{fig1}, we have plotted the slow-roll parameters
$\epsilon_1$, $\epsilon_2$, and $\epsilon_6$ in terms of the
e-folding number and the potential power with fixed values for the
rest of constants regardless of their units (if any).
Specifically, we have specified the potential power $n$ within the
range of $1.5$ to $4$ and $N$ within the range of $50$ to $70$.

Under the selected values mentioned, all the plots in
Fig.~\ref{fig1} illustrate that the slow-roll condition
$|\epsilon_i| \ll 1$ is maintained as it should be. Actually, the
behavior displayed in Fig.~\ref{fig1} provides support for the use
of the usual slow-roll approximations for the presented model. In
other words, the sufficiently small values obtained for the
slow-roll parameters confirm the internal consistency of the
approximation in the leading-order effects for the dynamics of
this modified gravity. Indeed, since the Chern-Simons correction
leaves the background Friedmann equations unchanged and
contributes only through the additional slow-roll parameter
$\epsilon_6$, its effect remains compatible with the slow-roll
evolution established here.
\begin{figure}[t!]
\centering
\includegraphics[width=0.5\textwidth]{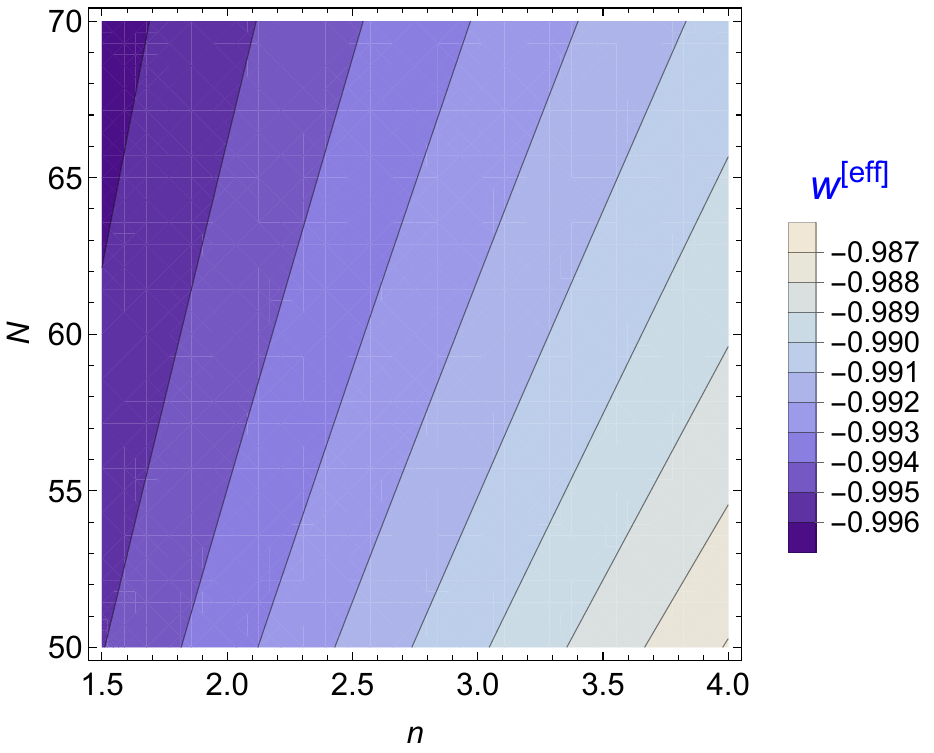}
 \caption{\label{fig3} [color online] The effective equation of state parameter during
 inflation has been plotted in terms of the e-folding number,
 $N$, and the potential power, $n$.}
\end{figure}

Now, let us derive the key inflationary observables, the scalar
spectral index, the tensor spectral index, and the
tensor-to-scalar ratio. These quantities have been evaluated in
terms of the slow-roll parameters as~\cite{Odintsov:2020sqy,
Venikoudis:2021oee}
\begin{align}
n_{\rm S}^{} &=1-\dfrac{4\epsilon_1+2\epsilon_2-2\epsilon_3+2\epsilon_4}{1-\epsilon_1},\label{nSFirst}\\
n_{\rm T}^{}&=-2\dfrac{\epsilon_1+\epsilon_6}{1-\epsilon_1},\label{nTFirst}\\
{\rm r}&=\Big|8(\epsilon_1+\epsilon_3)\sum_{i=L,R} \dfrac{F}{
Q_t(\lambda_i,\phi)}\Big|.\label{rFirst}
\end{align}
For the proposed model, utilizing relations \eqref{Qt},
\eqref{eps11p} and \eqref{eps66p}, these inflationary observables
become
\begin{eqnarray}
\label{lastns}n_{\rm S}^{}&\!\approx\! &\dfrac{\(1+N-2n\)}{N},\\
\label{lastnt}n_{\rm T}^{}&\!\approx\! &-\Bigg\lbrace\!
\dfrac{n^3\!\left(n\!+\!4N\right)\!\cos(\sqrt{{\rm
B}})\left[\kappa\Lambda v
\left(1\!+\!2\beta\right)\right]^2\!\left(\phi_1^2{\rm B}\right)^{n-2}}{2N\phi_1(1+\beta)} \nonumber\\
           && \times \dfrac{(n\!-\!1)\phi_1 \cos(\sqrt{{\rm B}})
           \!-\!\sqrt{\phi_1^2{\rm B}} \sin(\sqrt{{\rm B}})}{{\rm C}}+\dfrac{n}{2N}\!\Bigg\rbrace\!,\\
\label{lastr} r&\!\approx\! &\dfrac{18\, n^2
\left(1+\beta\right)}{\kappa^2 {\rm B}\,{\rm C}}.
\end{eqnarray}
\begin{figure}[t!]
\centering
\includegraphics[width=0.5\textwidth]{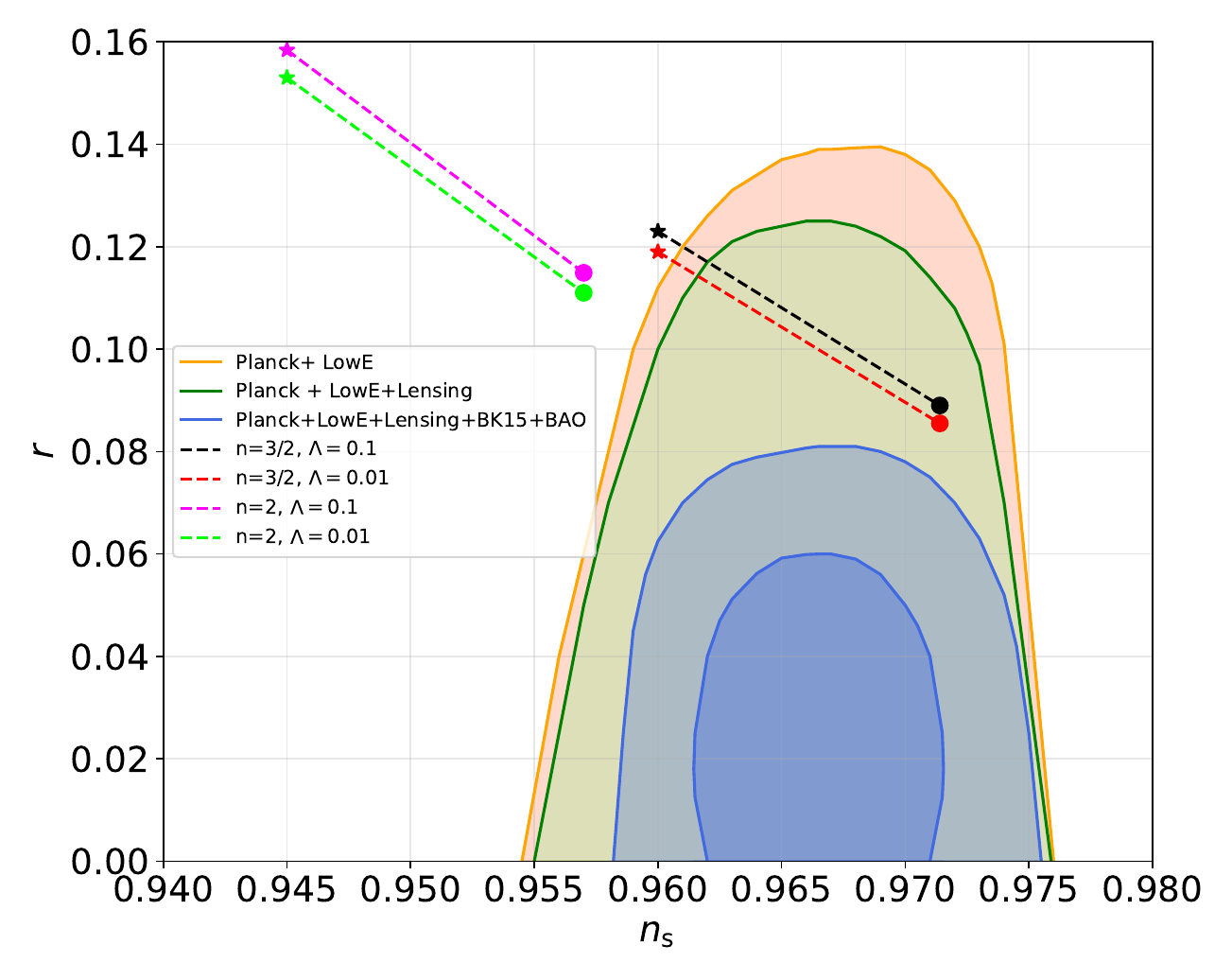}
 \caption{\label{fig4} [color online] Predictions of the linear form of $f(R,T)$ gravity with the Chern-Simons
 higher-curvature corrections for $n=1.5$ and $n=2$, each with two values $\Lambda=0.1$ and $\Lambda=0.01$,
 and in the range $N=[50,70]$, with the values $\beta=10^5$, $\phi_1=1/\kappa^2$, $v=1/(8 \pi)^2$
 and $\kappa^2=8 \pi$. The star indicates $N=50$ and the
 solid circle $N=70$, in regions for $n_{\rm S}^{}$ and $r$ at the pivot
 scale of $k=0.002~{\rm Mpc^{-1}}$ based on the Planck $2018$ data~\cite{Planck:2018jri}.}
\end{figure}
As is evident, while the scalar spectral index indicates
dependence only on $n$ and $N$, the tensor spectral index and the
tensor-to-scalar ratio additionally indicate more dependence on
the specific model parameters $\beta$, $v$, $\Lambda$, and
$\phi_1$. Hence, the contribution of the Chern-Simons correction
in $f(R, T)$ gravity extends the value of the tensor spectral
index and the tensor-to-scalar ratio. This enhancement can be
significance in unraveling the intricate dynamics of inflationary
phenomena, hence in this model, the characteristics observed
during inflation hinge on a set of key parameters $n$, $N$,
$\beta$, $v$, $\Lambda$, and $\phi_1$.

Before proceeding, we recall that within this context, the most
recent constraints established by the Planck collaboration
concerning the scalar spectral index and the tensor-to-scalar
ratio are~\cite{Planck:2018jri}
\begin{align}\label{planckdata}
&n_{\rm S}^{}= 0.9649\pm 0.0042\quad \nonumber\\
&{}\qquad {\rm at\!:}\ 68 \%\, {\rm CL, Planck\, TT,TE,EE+low\,E+lensing},\nonumber\\
& r<0.10 \qquad{\rm at\!:}\ 95 \%\, {\rm CL, Planck\,
TT+low\,E+lensing}.
\end{align}
However, by combining the Planck, BK$15$ and BAO data, additional
constraints have been imposed, resulting in the upper limit of $r$
being restricted to a specific value of
\begin{equation}\label{planckdata2}
r<0.056\qquad{\rm at}\!:\ 95 \%\, {\rm CL}.
\end{equation}

Now, in Fig.~\ref{fig2}, we have plotted the scalar spectral
index, $n_{\rm S}^{}$, the tensor spectral index, $n_{\rm T}^{}$,
and the tensor-to-scalar ratio, $r$, in terms of the potential
power and the e-folding number with fixed values for the rest of
constants regardless of their units (if any). This figure
indicates that, with the specified parameter configurations, the
outcomes for these inflationary observables are well compatible
with the Planck $2018$ data, especially for $1.5\leq n<2$ and
$60<N\leq 70$.  In this context, the tensor-to-scalar ratio is
somewhat elevated compared to the constraints obtained from the
joint data. As reported, the optimal values of the four free
parameters (i.e., $\beta$, $v$, $\Lambda$, and $\phi_1$) for the
tensor-to-scalar ratio have been adjusted to the most favorable
outcomes, and when these parameters are chosen to alternative
values (within reasonable limits), no improvement in the results
has been achieved up to our investigations.

Next, by inserting Eqs. \eqref{h2app}, \eqref{phidot}, \eqref{V1}
and \eqref{phiTo2} into the last expression of relation
\eqref{eos}, while neglecting the second-order of
$\dot{\phi}^2/V$, we obtain\footnote{Alternatively, it can be
obtained by using the slow-roll parameter $\epsilon_1$ defined in
relation~\eqref{defsr} into the middle expression of relation
\eqref{eos} while substituting relation~\eqref{eps11p}. }\
 the effective
equation of state parameter also as a function of the e-folding
number and the potential power as
\begin{align}\label{wEffect}
w^{[\rm eff]}\approx
-\dfrac{1}{3}\left(\dfrac{n+12N}{n+4N}\right).
\end{align}
Then, in Fig.~\ref{fig3}, we have plotted its behavior in terms of
$n$ and $N$. This parameter is also a significant quantity that
characterizes the nature of the cosmic fluid that causes the
expansion of the Universe during the inflationary epoch.
Considering relation~\eqref{wEffect}, when $n$ is negligible
compared to $N$, the effective equation of state parameter is very
close to $-1$ (as also shown below Eq.~\eqref{phidot}), indicating
a cosmic fluid consistent with the cosmological constant or the
vacuum energy, which is necessary for sustaining an inflationary
phase. This is confirmed in Fig.~\ref{fig3}, especially again for
$1.5\leq n<2$ and $60<N\leq 70$.

Accordingly, in Fig.~\ref{fig4}, by employing relations
\eqref{lastns} and \eqref{lastr}, we have also plotted the $(r,
n_{\rm S}^{})$ plane for two values $n=1.5$ and $n=2$, each with
two values $\Lambda=0.1$ and $\Lambda=0.01$, in the range
$N=[50,70]$, and with fixed values for the rest of constants
regardless of their units (if any). The obtained results show that
the model with $n=1.5$, despite providing predictions that are
better aligned with the Planck $2018$ data, is not in good
agreement with the joint data.

\subsubsection{\bf The Model without Chern-Simons Correction}

\begin{figure*}[t!]
\centering \subfigure[]{
\includegraphics[width=0.31\textwidth]{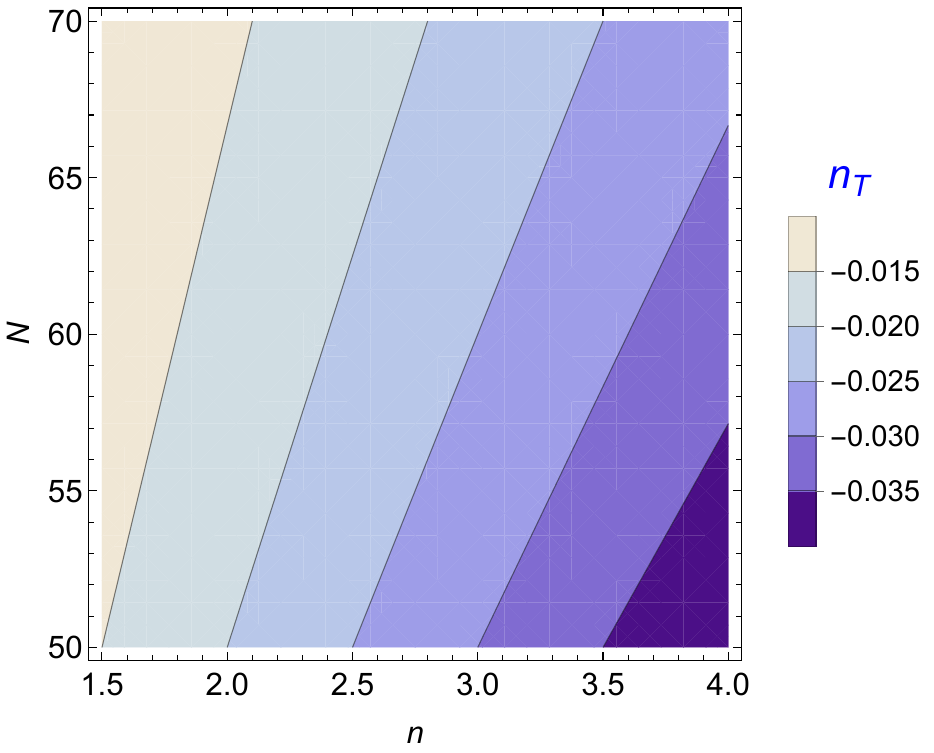}
\label{aa11} } \subfigure[] {
\includegraphics[width=0.32\textwidth]{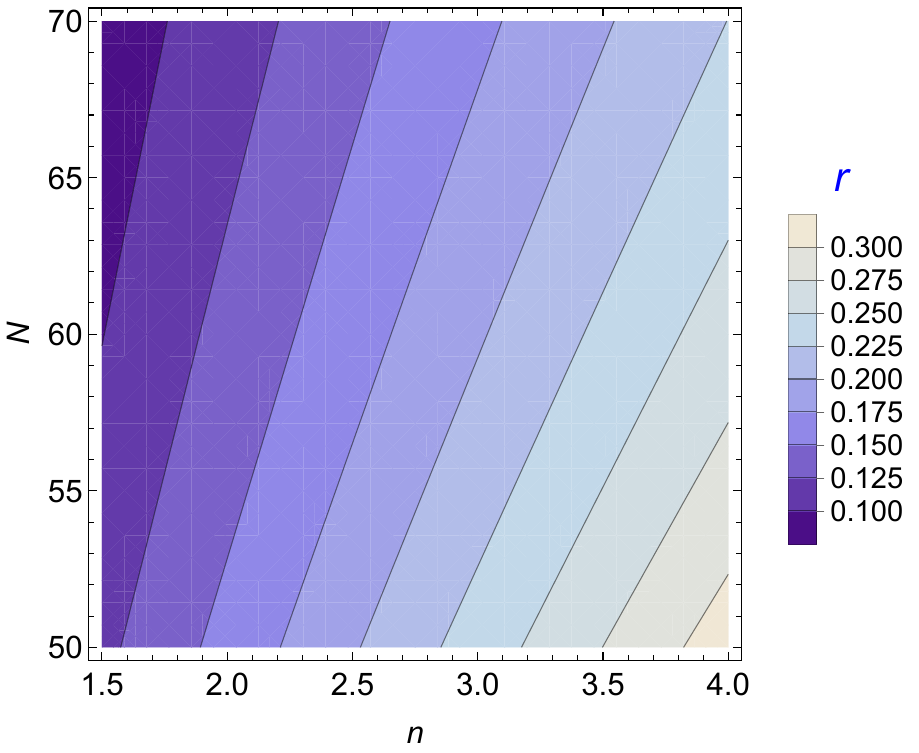}
\label{bb22} } \caption{\label{fig5} [color online] The
inflationary observables have been plotted in terms of the
e-folding number and the potential power in the range $N=[50,70]$
and $n=[1.5,4]$ in the absence of the Chern-Simons correction,
where (a) represents the range of the tensor spectral index,
$n_{\rm T}^{}$, and (b) the tensor-to-scalar ratio, $r$.}
\end{figure*}

In this scenario, by setting $\nu=0$, the model reduces to
$f(R,T)$ gravity plus the inflaton scalar field in the absence of
the Chern-Simons correction. Therefore, the slow-roll parameter
$\epsilon_6$ vanishes, however the rest of relevant relations
including the slow-roll parameters $\epsilon_1$ and $\epsilon_2$,
the scalar spectral index and the effective equation of state
parameter, relations~(\ref{eps11p}), (\ref{lastns})
and~(\ref{wEffect}), remain unchanged. Nevertheless, in this case,
the tensor spectral index and the tensor-to-scalar ratio,
relations~\eqref{nTFirst} and~\eqref{rFirst}, reduce to
\begin{align}
\label{nNEW}&n_{\rm T}^{}=-\dfrac{2\epsilon_1}{1-\epsilon_1},\\
\label{rNEW}&r=16\epsilon_1,
\end{align}
without any use of the slow-roll parameter $\epsilon_2$. Then, by
substituting relation~(\ref{eps11p}), the above relations become
\begin{eqnarray}
n_{\rm T}^{}&\!\approx\! &-\dfrac{n}{2N},\\
r&\!\approx\! &\dfrac{16n}{n+4N}.
\end{eqnarray}
As is obvious, the observables $n_{\rm T}^{}$ and $r$, like
$n_{\rm S}^{}$, depend only on the parameters $n$ and $N$; and
there is no~dependence on the model parameters $\beta$, $\Lambda$,
$\phi_1$ and $v$, and even $\kappa$. In Fig.~\ref{fig5}, we have
plotted these two quantities in the ranges $n=[1.5,4]$ and
$N=[50,70]$. The results obtained demonstrate that these
inflationary observables are in good agreement with the
constraints of Planck $2018$ data. This consistency aligns with
the results reported in Ref.~\cite{Gamonal:2020itt}.
Interestingly, the findings show very little deviation from the
predictions of the linear form of $f(R,T)$ gravity with the
power-law potential in the presence of the Chern-Simons
correction. Nevertheless, we have examined the difference between
the two cases, in more detail, through calculating the ratio
$n_{\rm T}^{}$, as well as $r$, in the presence and absence of the
Chern-Simons correction. The results show that the difference in
the corresponding ratio for both $n_{\rm T}$ and $r$ with $1$ is
approximately of the order of ${\cal O}(10^{-3})$.

\subsubsection{\bf The Model
              without the Linear Form of $f(R,T)$ Gravity}

In this situation, by setting $\beta =0$ (in effect by eliminating
the trace of the energy-momentum tensor), the model reduces to GR
plus the inflaton scalar field in the presence of the Chern-Simons
correction. Also, in this case, the slow-roll parameters
$\epsilon_1$ and $\epsilon_2$, the scalar spectral index and the
effective equation of state parameter, relations~(\ref{eps11p}),
(\ref{lastns}) and~(\ref{wEffect}), remain unchanged. To derive
the final results for the slow-roll parameter $\epsilon_6$ and the
inflationary observables $n_{\rm T}^{}$ and $r$, it is sufficient
to set $\beta=0$ in relations~\eqref{eps66p}, \eqref{lastnt}
and~\eqref{lastr}, respectively. We then plotted the new relations
for $n_{\rm T}^{}$ and $r$, and the results yield almost the same
results as for the linear form of $f(R,T)$ gravity in
Fig.~\ref{fig2} with the same values $\phi_1=1/\kappa^2$,
$v=1/(8\pi)^2$ and $\kappa^2=8 \pi$ in the range $n=[1.5,4]$ and
$N=[50,70]$, but with the choice of
$\Lambda=10^{-4}$.\footnote{Hence, we did~not include these plots
within the text.}\
 Furthermore, with these values, the slow-roll
parameter $\epsilon_6$ varies in the range $5\times
10^{-6}<\epsilon_6<4\times 10^{-5}$, as it should be much less
than $1$. As is clear, the model without the linear term when
incorporating the Chern-Simons correction produces predictions
compatible with the Planck $2018$ data that are comparable to the
results of the linear model with the Chern-Simons correction, but
by adjusting the parameter $\Lambda$. However, none of the models
provide results compatible with the joint Planck, BK$15$ and BAO
data.

\subsection{A Hilltop Potential}
As the second case, we choose a type of hilltop potential defined
as~\cite{Venikoudis:2021oee}
\begin{equation}\label{V2}
V(\phi)=\frac{\phi^4}{\left(1+\kappa^2\gamma \phi^2\right)^2},
\end{equation}
where the parameter $\gamma$ is a dimensionless constant.

\subsubsection{\bf The Model with Chern-Simons Exponential Coupling Function}

Then, we investigate the Chern-Simons exponential coupling
function to be~\cite{Venikoudis:2021oee}
\begin{equation}\label{nu2}
\nu(\phi)=\exp\left(-\kappa \phi\right).
\end{equation}

Now, for this case, we can achieve the slow-roll parameters
\eqref{eps1} and \eqref{eps6} as
\begin{align}
\label{eps12} &\epsilon_1 \approx \dfrac{8}{\kappa^2(1+\beta) \phi^2\left(1+\kappa^2\gamma \phi^2\right)^2}, \\
\label{eps22}&\epsilon_2 \approx \dfrac{12(1-\kappa^2\gamma \phi^2)}{\kappa^2 \phi^2 (1+\kappa^2\gamma \phi^2)^2}, \\
\label{eps62}&\epsilon_6 \approx \frac{1}{1+\beta}\left[\frac{8
\kappa^2 \left(1+2\beta\right)
\phi^2 \exp\left(-\kappa \phi\right)}{1+\kappa^2\gamma \phi^2}\right]^2  \nonumber\\
&\hspace*{-0.25cm}\times\frac{3- \kappa^3\gamma \phi^3-3
\kappa^2\gamma \phi^2- \kappa\phi}{9(1\!+\!\beta)^2_{}\!
\left(1\!+\!\kappa^2\gamma \phi^2\right)^6\!/4\!-\!\left[4\kappa^3
(1\!+\!2\beta) \phi^3 \exp\left(-\kappa \phi\right)\right]^2}.
\end{align}

Using the condition of the end of inflation for this case, the
value of the inflaton scalar field at the end of inflation is
obtained as
\begin{equation}\label{phiend2}
\phi_{\rm end}^2\approx
\frac{(1+\beta)^{1/3}\left[D(1+\beta)^{-1/3}-1\right]^2}{3\kappa^2
\gamma D},
\end{equation}
where
\begin{equation}
D\equiv \left[ 6\sqrt{6
\gamma(54\gamma+1+\beta)}+108\gamma+1+\beta\right]^{1/3}.
\end{equation}
In addition, by inserting the potential function \eqref{V2} into
relation \eqref{Neq}, the e-folding number reads
\begin{equation}
N\approx\frac{\kappa^2\left(1+\beta\right)}{16}\Big[
\kappa^2\gamma\left(\phi^4-\phi_{\rm end}^4\right)
+2\left(\phi^2-\phi_{\rm end}^2\right)\Big],
\end{equation}
and then from this relation, we can achieve the inflaton scalar
field at the first horizon crossing as
\begin{equation}\label{phii}
\phi^2\approx \dfrac{\!-\!1\!+\!\Big[\kappa^2 \gamma\phi_{\rm
end}^2 (\kappa^2 \gamma \phi_{\rm end}^2+2) +\frac{16\gamma\,
N}{1+\beta}+1\Big]^{1/2}}{\kappa^2 \gamma }.
\end{equation}
Thereafter, by substituting relation \eqref{phii}, while using
relation \eqref{phiend2}, into relations \eqref{eps12},
\eqref{eps22} and \eqref{eps62}, the slow-roll parameters can be
extracted as functions of the parameters $N$, $\beta$ and
$\gamma$.

\begin{figure*}[t!]
\centering \subfigure[]{
\includegraphics[width=0.32\textwidth]{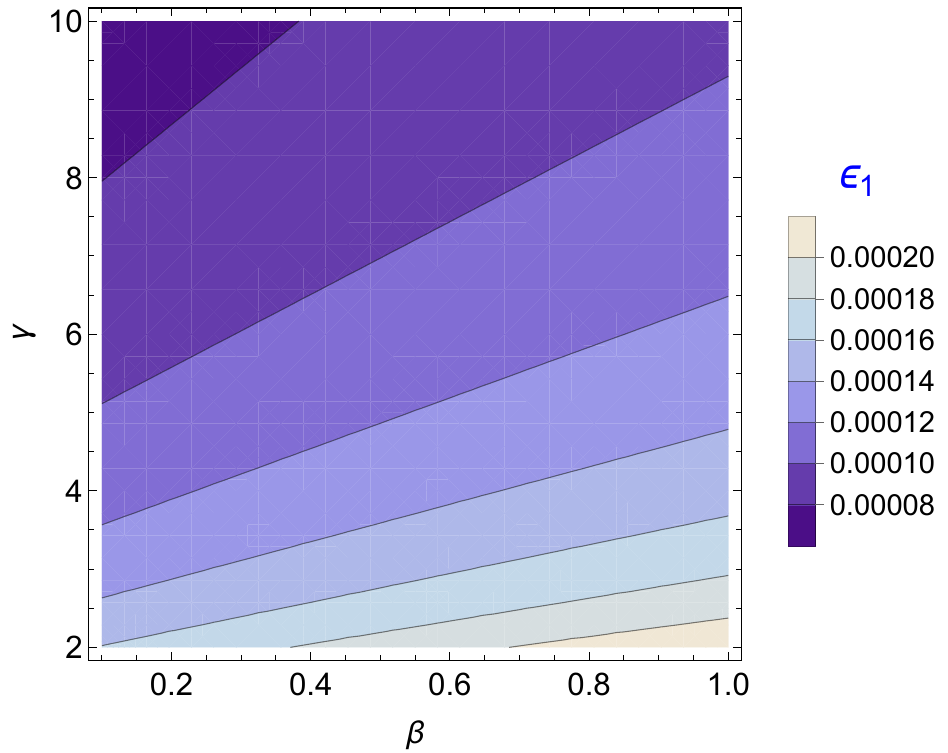}
\label{a1} } \subfigure[] {
\includegraphics[width=0.31\textwidth]{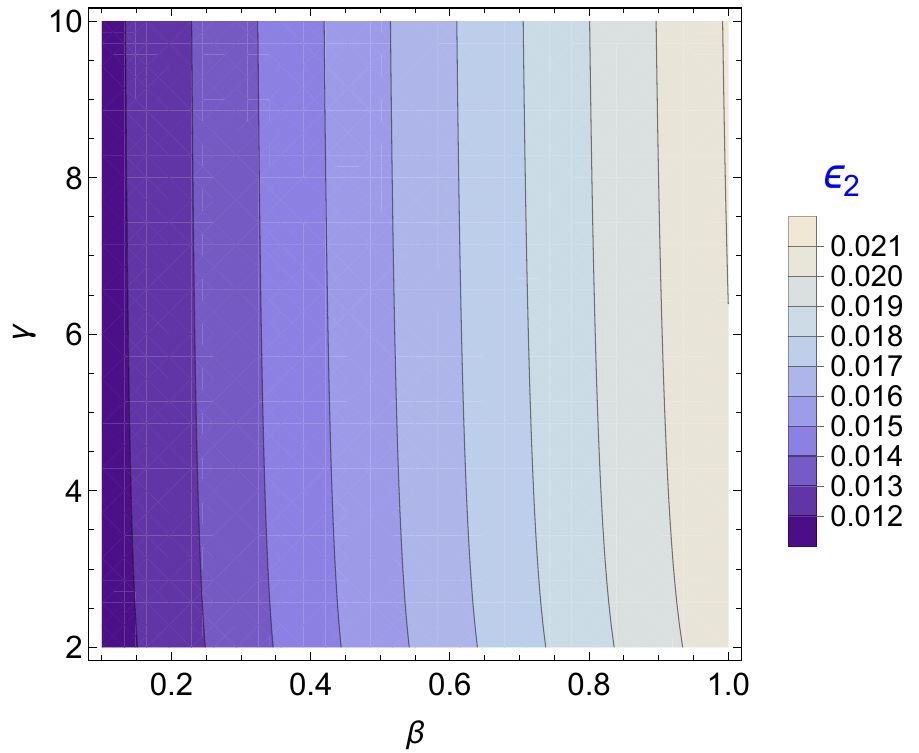}
\label{b1} } \subfigure[] {
\includegraphics[width=0.32\textwidth]{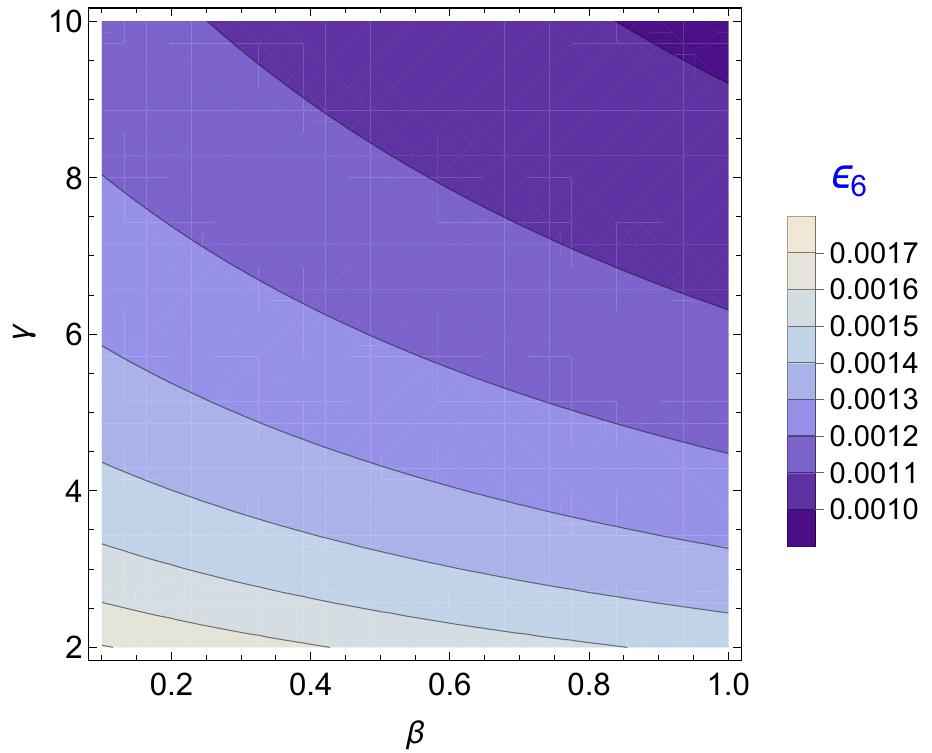}
\label{c1} }
 \caption{\label{fig6} [color online] The slow-roll parameters have been plotted
 in terms of the parameters $\beta$ and $\gamma$, for $N=70$ with the value
 $\kappa^2=8\pi$ in the range $\beta=[0.1,1]$ and $\gamma=[2,10]$,
 where (a) shows the range of parameters $\epsilon_1$, (b)
 $\epsilon_2$, and (c) $\epsilon_6$.}
\end{figure*}

In Fig.~\ref{fig6}, we have plotted these slow-roll parameters for
$N=70$ in terms of the model parameters $\beta$ and $\gamma$ in
the range $\beta=[0.1, 1]$ and $\gamma=[2, 10]$. The results are
in good agreement with the condition $|\epsilon_i|\ll 1$, as they
should. Again, these results confirm the internal consistency of
the approximation in the leading-order effects for the dynamics of
the presented modified gravity.

\begin{figure*}[t!]
\centering \subfigure[]{
\includegraphics[width=0.31\textwidth]{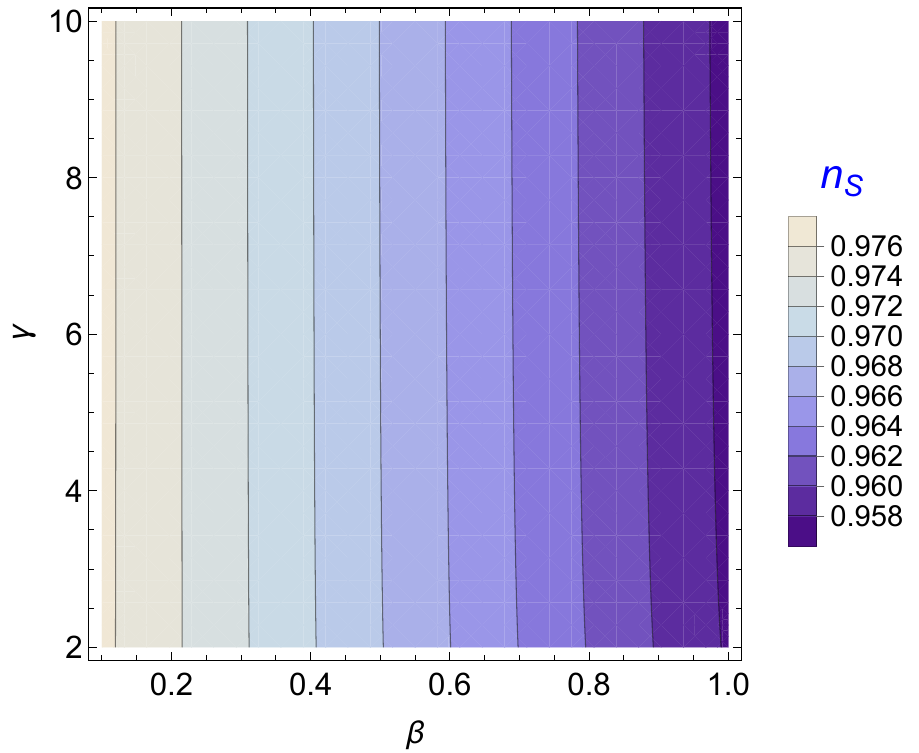}
\label{a1} } \subfigure[] {
\includegraphics[width=0.32\textwidth]{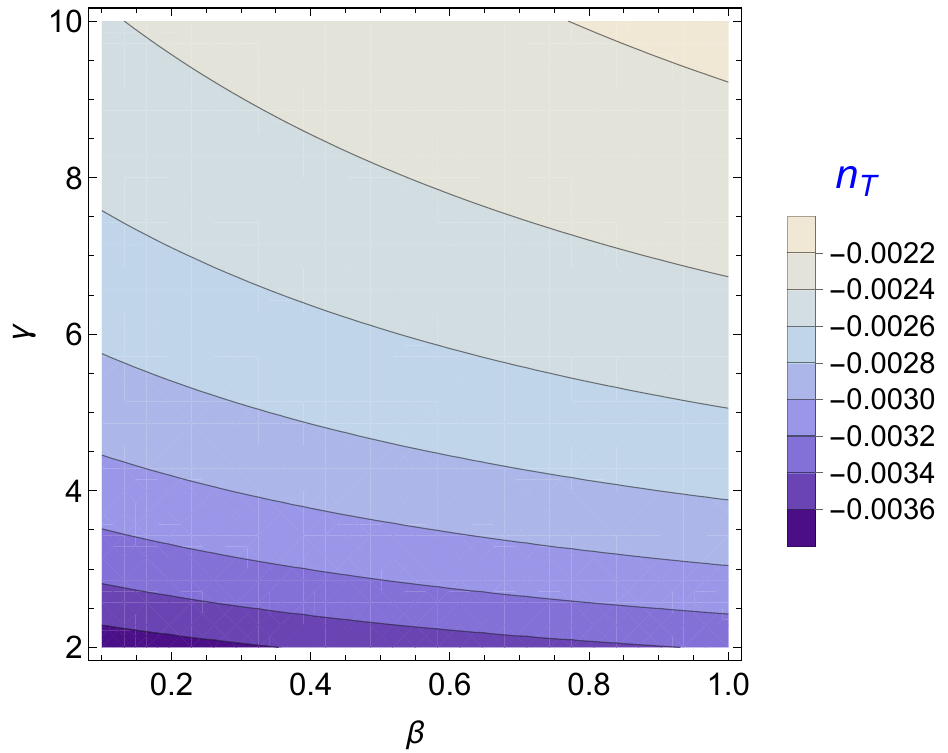}
\label{b1} } \subfigure[] {
\includegraphics[width=0.31\textwidth]{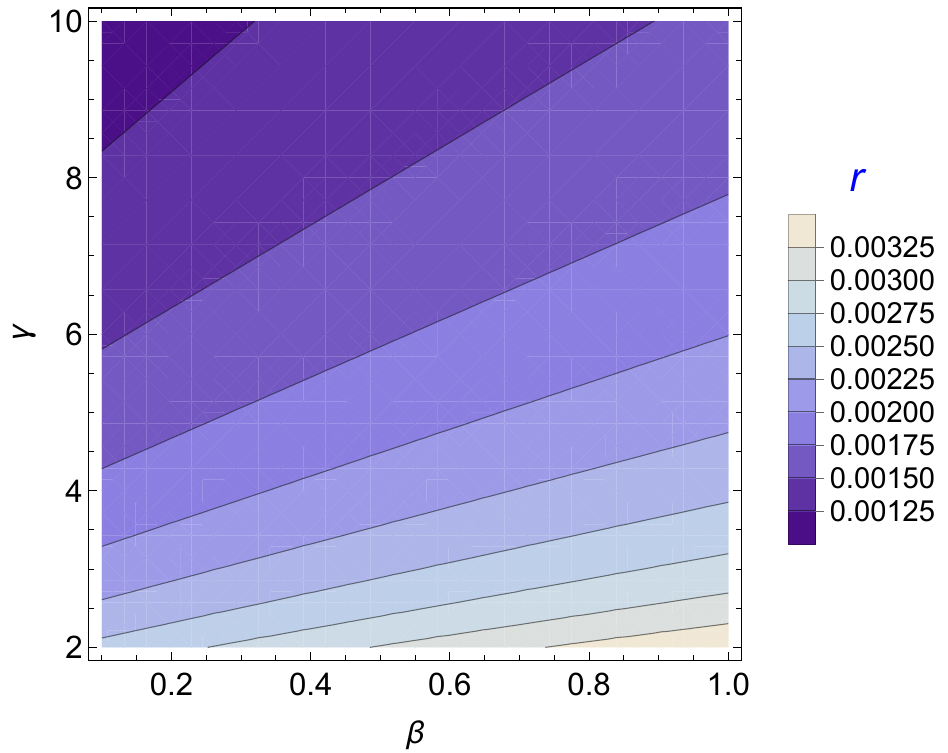}
\label{c1} }
 \caption{\label{fig7} [color online] The inflationary observables have been plotted in terms
 of the parameters $\beta$ and $\gamma$, for $N=70$ with the value
 $\kappa^2=8\pi$ in the range $\beta=[0.1,1]$ and $\gamma=[2,10]$,
 where (a) shows the range of the scalar spectral index, $n_{\rm S}^{}$, (b) the tensor
 spectral index, $n_{\rm T}^{}$, and (c) the tensor-to-scalar ratio, $r$.}
\end{figure*}

At this stage, by applying relations \eqref{eps12}, \eqref{eps22}
and \eqref{eps62}, while using relations \eqref{phiend2} and
\eqref{phii}, into relations \eqref{nSFirst}, \eqref{nTFirst} and
\eqref{rFirst}, we can obtain the inflationary observables $n_{\rm
S}^{}$, $n_{\rm T}^{}$ and $r$ as functions of the model
parameters $\beta$ and $\gamma$ and the e-folding number. Then, in
Fig.~\ref{fig7}, we have plotted these inflationary observables
for $N=70$ in the range $\beta=[0.1, 1]$ and $\gamma=[2, 10]$. The
results obtained are in good agreement with both the Planck $2018$
data, \eqref{planckdata}, and the joint Planck, BK$15$ and BAO
data, \eqref{planckdata2}. Accordingly, we conclude that the
proposed model with the case of the Chern-Simons exponential
coupling function and potential \eqref{V2} can impose a stricter
limit on the value of the tensor-to-scalar ratio.

\begin{figure}[t!]
\centering
\includegraphics[width=0.5\textwidth]{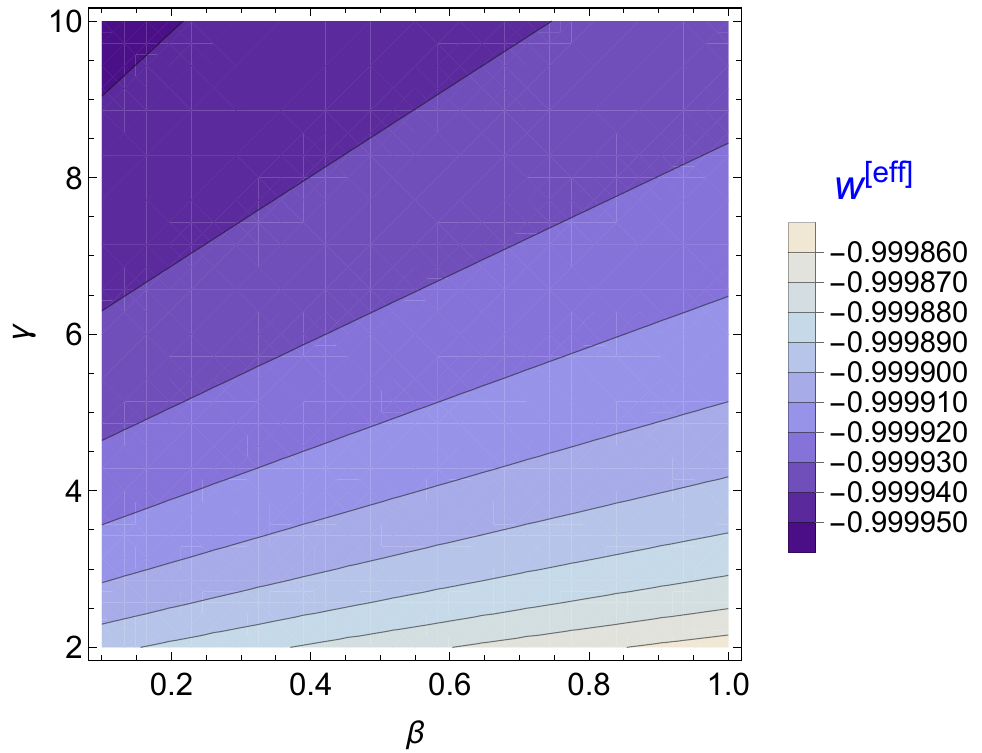}
 \caption{  [color online] The effective equation of state parameter during inflation
 has been plotted in terms of the parameters $\beta$ and $\gamma$, for $N=70$ with the value
 $\kappa^2=8\pi$ in the range $\beta=[0.1,1]$ and $\gamma=[2,10]$.}
 \label{fig8}
\end{figure}

Also, by using the slow-roll parameter $\epsilon_1$ defined in
relation~\eqref{defsr} into the middle expression of relation
\eqref{eos}, while substituting relation~\eqref{eps12}, we obtain
the effective equation of state parameter for this case as
\begin{equation}\label{wEffHilltop}
w^{[{\rm eff}]}\approx -1+\frac{16}{3 \kappa^2 (1+\beta) \phi^2
(1+ \kappa^2 \gamma\phi^2)^2}.
\end{equation}
Then, using relations \eqref{phiend2} and \eqref{phii}, we can
derive it as a function of the e-folding number and the parameters
$\beta$ and $\gamma$. Thereafter, in Fig.~\ref{fig8}, we have
plotted this parameter for $N=70$ in the range $\beta=[0.1, 1]$
and $\gamma=[2, 10]$. The figure indicates that the model
prediction for this parameter is in good agreement with its
expected value during the inflationary era.

\begin{figure}[t!]
\centering
\includegraphics[width=0.5\textwidth]{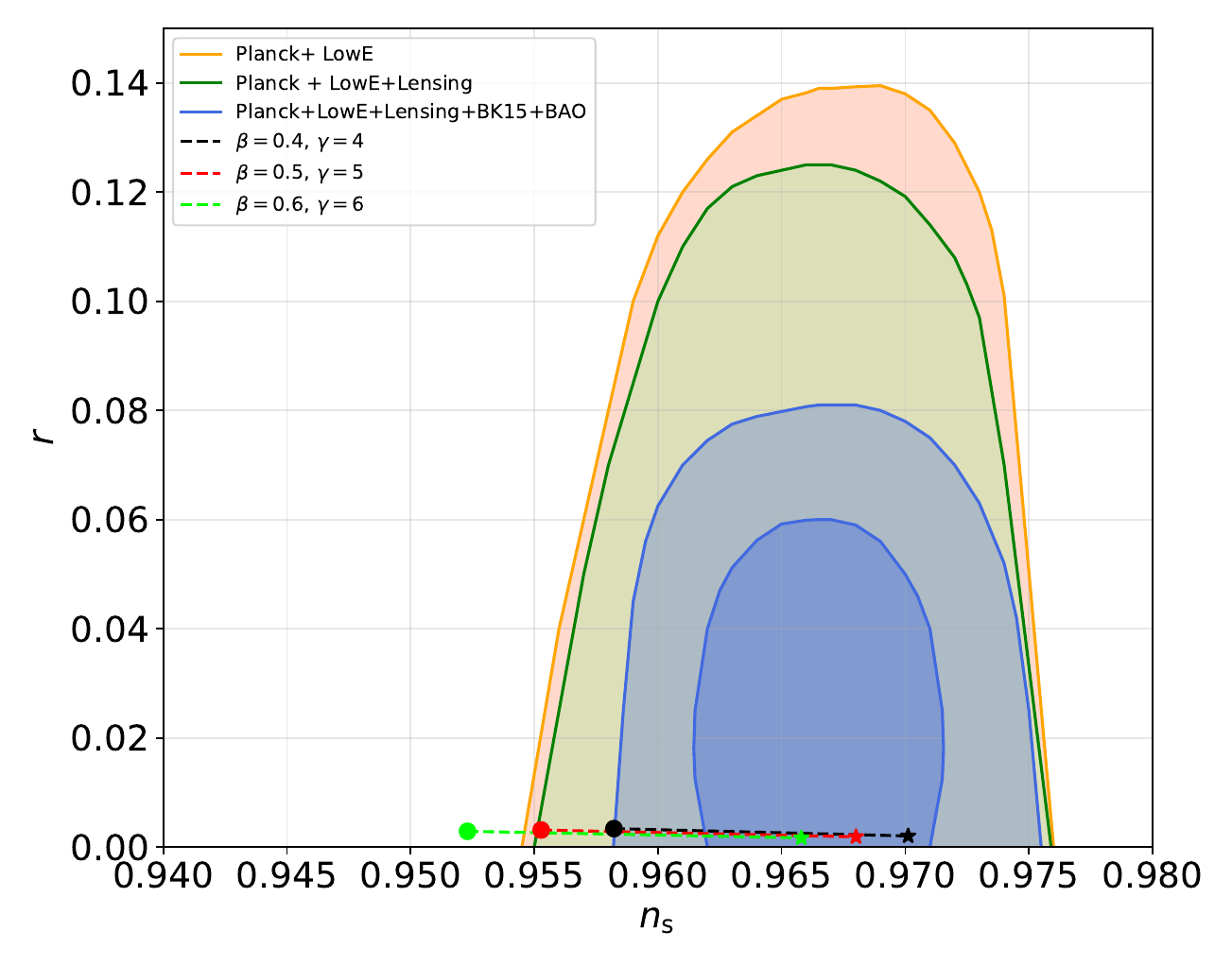}
 \caption{ [color online] Predictions of the linear form of $f(R,T)$ gravity with the Chern-Simons
 higher-curvature corrections for $(\beta=0.4,\gamma=4)$, $(\beta=0.5,\gamma=5)$ and $(\beta=0.6,\gamma=6)$,
 with the value $\kappa^2=8 \pi$ and in the range $N=[50,70]$
 for the $(r, n_{\rm S}^{})$ plane at the pivot scale of $k=0.002~{\rm
 Mpc^{-1}}$. The star represents $N=50$ and the solid circle $N=70$, based on the Planck 2018 data~\cite{Planck:2018jri}.}
 \label{fig9}
\end{figure}

Moreover, in Fig. \ref{fig9}, we have once again plotted the $(r,
n_{\rm S}^{})$ plane, however for the values of
$(\beta=0.4,\gamma=4)$, $(\beta=0.5,\gamma=5)$ and
$(\beta=0.6,\gamma=6)$ in the range $N=[50,70]$. The results also
confirm that the proposed model in this case produces predictions
that are aligned with both observational data \eqref{planckdata}
and \eqref{planckdata2}.

\subsubsection{\bf The Model without Chern-Simons Correction}

\begin{figure}[t!]
\centering
\includegraphics[width=0.45\textwidth]{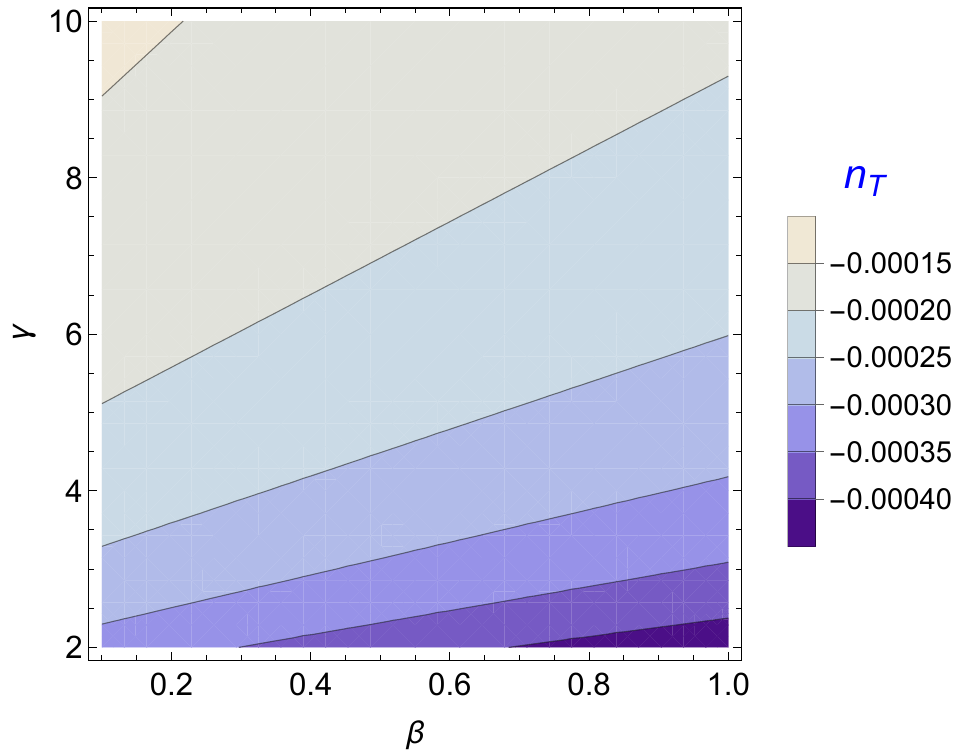}
 \caption{\label{fig10} [color online] The
inflationary observable $n_{\rm T}^{}$ has been plotted in terms
of the parameters $\beta$ and $\gamma$, for $N=70$ with the value
$\kappa^2=8\pi$ in the range $\beta=[0.1,1]$ and $\gamma=[2,10]$
in the absence of Chern-Simons correction.}
\end{figure}

In this case too, by setting $\nu=0$, as in the case $A2$, the
model reduces to $f(R,T)$ gravity plus the inflaton scalar field
in the absence of the Chern-Simons correction. Hence, the
slow-roll parameter $\epsilon_6$ vanishes, however the rest of
relevant relations including the slow-roll parameters $\epsilon_1$
and $\epsilon_2$, the scalar spectral index and the effective
equation of state parameter remain unchanged. Again, in this case,
the tensor spectral index and the tensor-to-scalar ratio,
relations~\eqref{nTFirst} and~\eqref{rFirst}, reduce to
relations~\eqref{nNEW} and~\eqref{rNEW} without any use of the
slow-roll parameter $\epsilon_2$. Then, by applying
relations~\eqref{eps12} and \eqref{eps22}, while utilizing
relations~\eqref{phiend2} and~\eqref{phii}, into
relations~\eqref{nNEW} and~\eqref{rNEW}, we obtain the
inflationary observables $n_{\rm T}^{}$ and $r$ as functions of
the model parameters $\beta$ and $\gamma$ and the e-folding
number. Next, in Fig.~\ref{fig10}, we have plotted the
inflationary observable $n_{\rm T}^{}$ for $N=70$ and
$\kappa^2=8\pi$ in the range $\beta=[0.1,1]$ and $\gamma=[2,10]$.
The result indicates smaller values than the corresponding result
of the previous subsection in Fig.~\ref{fig7}(b). Likewise, we
plotted the inflationary observable $r$ with the same values, and
the result is the same as the linear form of $f(R,T)$ gravity in
Fig.~\ref{fig7}(c),\footnote{Hence, we did~not include this plot
within the text.}\
 which is in good agreement with
both the Planck $2018$ data and the joint Planck, BAO and BK$15$
data. The calculated ratio $r$ in the presence and absence of the
Chern-Simons correction shows that its difference with $1$ is
approximately of the order of ${\cal O}(10^{-8})$.

\subsubsection{\bf The Model
              without the Linear Form of $f(R,T)$ Gravity}

\begin{figure*}[t!]
\centering \subfigure[]{
\includegraphics[width=0.32\textwidth]{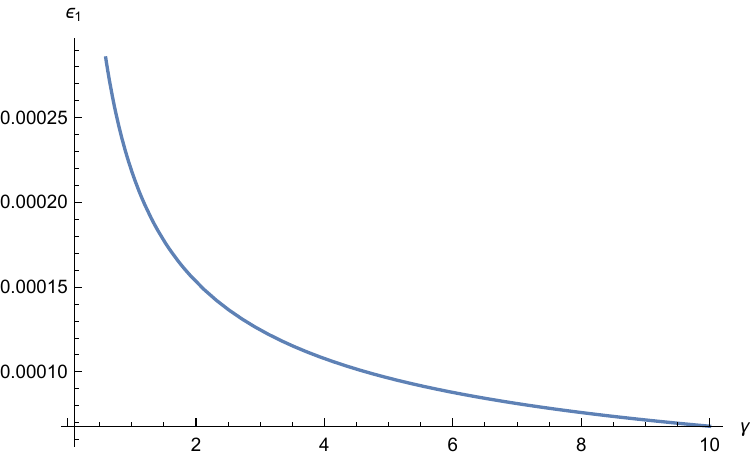}
\label{a1} } \subfigure[] {
\includegraphics[width=0.31\textwidth]{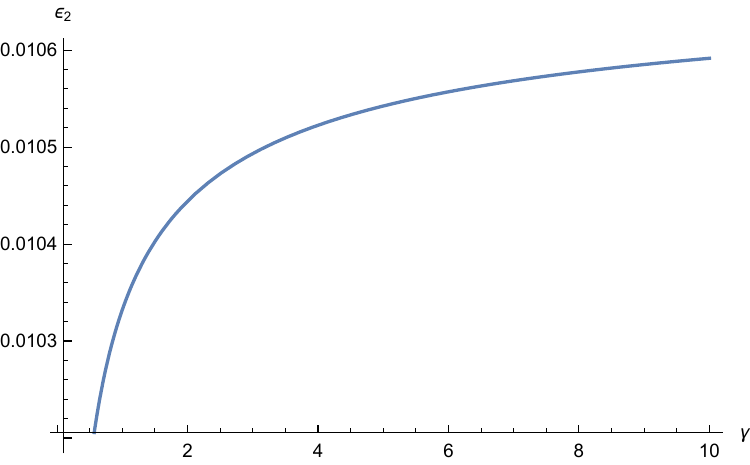}
\label{b1} } \subfigure[] {
\includegraphics[width=0.32\textwidth]{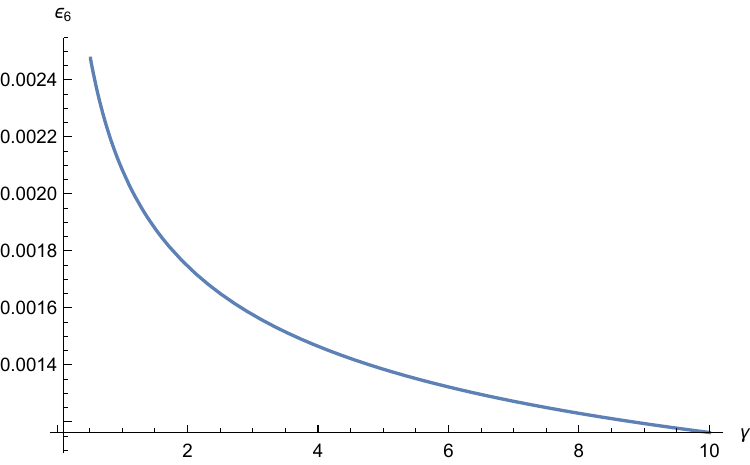}
\label{c1} }
 \caption{\label{fig11} The slow-roll parameters have been plotted
 in terms of the parameter $\gamma$, for $N=70$ with the value $\kappa^2=8 \pi$ in the
range $\gamma=[0.1, 10]$,
 where (a) shows the range of parameters $\epsilon_1$, (b)
 $\epsilon_2$, and (c) $\epsilon_6$.}
\end{figure*}

\begin{figure*}[t!]
\centering \subfigure[]{
\includegraphics[width=0.31\textwidth]{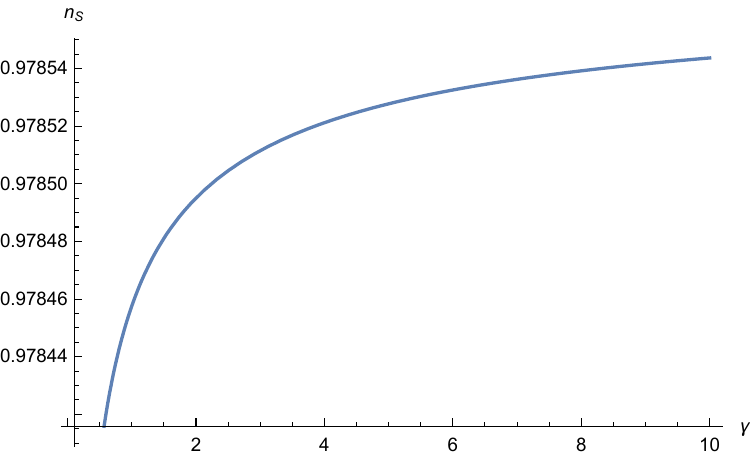}
\label{a1} } \subfigure[] {
\includegraphics[width=0.32\textwidth]{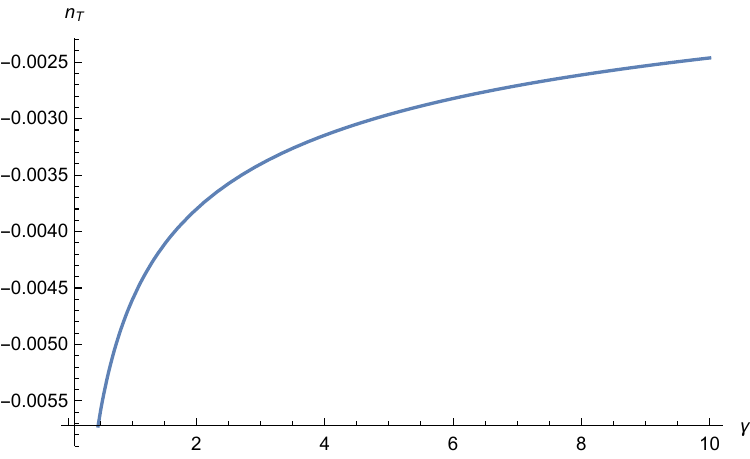}
\label{b1} } \subfigure[] {
\includegraphics[width=0.31\textwidth]{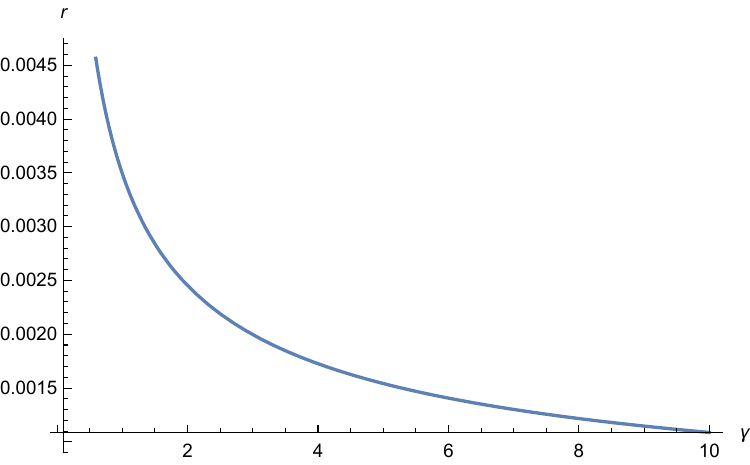}
\label{c1} }
 \caption{\label{fig12} The inflationary observables have been plotted in terms
 of the parameter $\gamma$, for $N=70$ with the value $\kappa^2=8 \pi$ in the
range $\gamma=[0.1, 10]$, where (a)
 shows the range of the scalar spectral index, $n_{\rm S}^{}$, (b) the tensor
 spectral index, $n_{\rm T}^{}$, and (c) the tensor-to-scalar ratio, $r$.}
\end{figure*}

\begin{figure}[t!]
\centering
\includegraphics[width=0.5\textwidth]{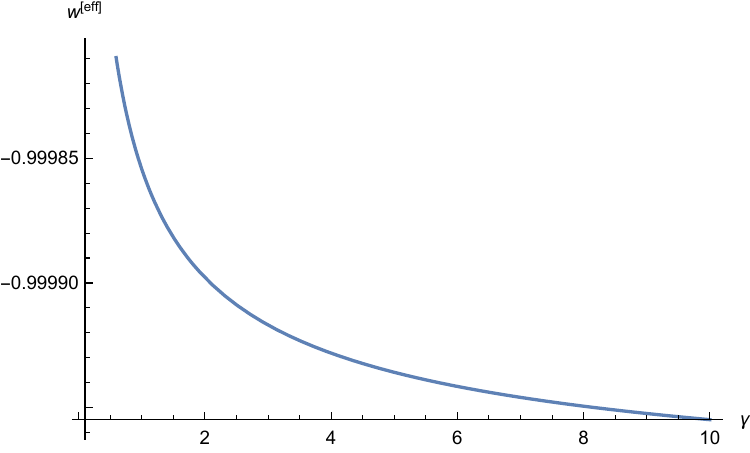}
 \caption{ The effective equation of state parameter during inflation
 has been plotted in terms of the parameter $\gamma$, for $N=70$ with the value $\kappa^2=8 \pi$ in the
range $\gamma=[0.1, 10]$.}
 \label{fig13}
\end{figure}

In this situation too, by setting $\beta =0$ (in effect by
eliminating the trace of the energy-momentum tensor), as in the
case $A3$, the model reduces to GR plus the inflaton scalar field
in the presence of the Chern-Simons correction. In this case, the
appearance of relation~\eqref{eps22} remains unchanged, although
the inflaton scalar field mentioned in it changes in this case.
Then, we set $\beta=0$ in relations~\eqref{eps12}, \eqref{eps62},
\eqref{phiend2}, \eqref{phii} and~\eqref{wEffHilltop} to get the
corresponding new relations for them. Next, to derive the final
results for the slow-roll parameters $\epsilon_1$, $\epsilon_2$
and $\epsilon_6$, the inflationary observables $n_{\rm S}^{}$,
$n_{\rm T}^{}$ and $r$, and the effective equation of state
parameter, it is sufficient to insert relations~\eqref{phiend2}
and~\eqref{phii} into relation~\eqref{eps22} and into
relations~\eqref{eps12}, \eqref{eps62} and~\eqref{wEffHilltop}.
Thereafter, in Figs.~\ref{fig11}, \ref{fig12} and \ref{fig13}, we
have plotted the corresponding slow-roll parameters, inflationary
observables and effective equation of state parameter for $N=70$
with the value $\kappa^2=8 \pi$ in terms of $\gamma=[0.1, 10]$.
All the plots in Fig.~\ref{fig11} illustrate that the slow-roll
condition $|\epsilon_i| \ll 1$ confirms the use of the usual
slow-roll approximations for the presented model. Also, the
results obtained in Figs.~\ref{fig12} and \ref{fig13} show that
all the inflationary observables are compatible with both the
Planck $2018$ data and the joint Planck, BAO and BK$15$ data,
which are in agreement with the results of the model with the
linear form of $f(R,T)$ gravity. However, the $n_{\rm S}^{}$
result is slightly better in the latter case. Furthermore, the
calculated ratio $r$ in the presence (e.g., with $\beta=0.1$) and
absence of the linear term, in the range $\gamma=[2, 10]$, shows
that its difference with $1$ is approximately of the order of
${\cal O}(10^{-2})$.


\section{\bf
              The Model with a Non-Linear Form of $f(R,T)$ Gravity}

To further explore the motivation for considering the linear form
of $f(R,T)$ gravity, we also want to examine a non-linear form of
it in this section. However, there are many options for such
forms, hence as an example, we consider the $R\, T$ form as a
non-linear one. More explicitly, in action~\eqref{action}, we
investigate $f(R,T)=R+\alpha\kappa^4 R\, T$ gravity, where $\alpha
$ is a dimensionless constant, plus the inflaton scalar field in
the presence of the Chern-Simons correction. Such a non-linear
form has already been considered in Ref.~\cite{Sarkar} in the
absence of the Chern-Simons correction with other potentials.

Now, by varying this new action with respect to the metric and the
inflaton scalar field, and then using the FLRW metric, we obtain
the new modified Friedmann, Raychadhuri and Klein-Gordon
equations, respectively, as
\begin{eqnarray}\label{NewField1}
3H^2=&&\dfrac{\kappa^2}{
2}\Big[\left(1+18\alpha\kappa^2H^2+12\alpha\kappa^2\dot{H}\right)\dot{\phi}^2\cr
&& \!\!\!\!\!\!+2V\left(1+12\alpha\kappa^2 H^2\right)\Big],
\end{eqnarray}
\begin{eqnarray}\label{NewField2}
&3&\!\! H^2+3\dot{H}=\cr
&-&\kappa^2\Bigg[\left(1+9\alpha\kappa^2H^2+6\alpha\kappa^2\dot{H}-6\alpha\kappa^2V''\right)\dot{\phi}^2\cr
&&{}\,\,\,\,\,\,\,\,
-V\left(1+12\alpha\kappa^2H^2+12\alpha\kappa^2\dot{H}\right)\Bigg]\cr
&+& \kappa^4\left(-3\alpha H\dot{\phi}\ddot{\phi}+6\alpha
H\dot{\phi}V'\!-3\alpha\ddot{\phi}^2\!-3\alpha\dot{\phi}\dddot{\phi}\!+6\alpha\ddot{\phi}V'\right),\cr
&
\end{eqnarray}
\begin{eqnarray}\label{NewField3}
&&\ddot{\phi}\left(1+12\alpha\kappa^2H^2+6\alpha\kappa^2\dot{H}\right)\cr
&&+V'\left(1+24\alpha\kappa^2H^2+12\alpha\kappa^2\dot{H}\right)\cr
&&+3H\dot{\phi}\left(1+10\alpha\kappa^2H^2+8\alpha\kappa^2\dot{H}\right)\cr
&&+6\alpha\kappa^2\dot{\phi}\left(H^3+3H\dot{H}+\ddot{H}\right)=0.
\end{eqnarray}
Next, for inflationary epoch of the Universe, in addition to the
two usual slow-roll conditions, we require to add the extra
slow-roll conditions~\cite{Sarkar}
\begin{equation}\label{NewConditions}
\mid\!\dddot{\phi}\!\mid\ll\mid\! H\ddot{\phi}\!\mid\ll\mid\!
H^2\dot{\phi}\!\mid\quad {\rm and}\quad
\mid\!\ddot{H}\!\mid\ll\mid\! H\dot{H}\!\mid\ll\mid\! H^3\!\mid.
\end{equation}
Then, by applying all the slow-roll conditions in
Eqs.~\eqref{NewField1}, \eqref{NewField2} and~\eqref{NewField3},
we obtain
\begin{equation}\label{NewField11}
3H^2\approx \frac{\kappa^2 V}{1-4\alpha\kappa^4 V},
\end{equation}
\begin{equation}\label{NewField22}
\dot{H}\approx
-\frac{\kappa^2\dot{\phi}^2}{2\left(1-4\alpha\kappa^4 V\right)^2},
\end{equation}
\begin{equation}\label{NewField33}
3H\dot{\phi}+V'\left(1+4\alpha\kappa^4 V\right)\approx 0.
\end{equation}
On the other hand, using Eqs.~\eqref{NewField11}
and~\eqref{NewField22}, we get the useful equation
\begin{equation}\label{NewPhiDot}
3H\dot{\phi}\approx -V'.
\end{equation}
Accordingly, Eq.~\eqref{NewField33} dictates that its last term
can be ignored compared with its first two terms.

Thereafter, utilizing these new field equations into the
definitions~\eqref{defsr}, we obtain
\begin{equation}\label{NewSRe1,2}
\epsilon_1\approx
\frac{1}{2\kappa^2}\left(\frac{V'}{V}\right)^2\quad{\rm and}\quad
\epsilon_2\approx
\big|\frac{1}{\kappa^2}\frac{V''}{V}-4\alpha\kappa^2 V''\big|.
\end{equation}
However, in this case, we have $F=\kappa^{-2}(1+\alpha\kappa^4
T)$. To proceed, for simplicity, we consider the trace of the
energy-momentum to be the corresponding one for the inflaton
scalar field, i.e.
\begin{equation}\label{EMofSF}
T^{[\phi]}=\dot{\phi}^2-4V\approx -4V.
\end{equation}
Hence, $F\approx\kappa^{-2}(1-4\alpha\kappa^4 V)$, $E\approx
24\alpha^2\kappa^4 V'^2+F$, and in turn after some approximations,
we obtain
\begin{equation}\label{NewSRe3,4,5}
\epsilon_5=\epsilon_3\approx
2\alpha\kappa^2\left(\frac{V'^2}{V}\right) \approx\epsilon_4,
\end{equation}
\begin{eqnarray}\label{NewSRe6}
\epsilon_6\approx
 &&\, 4\kappa^2V'^2\left(1-4\alpha\kappa^4 V\right)\cr
 &&\times\dfrac{ \left[9\alpha\left(1-4\alpha\kappa^4V\right)
 +V'\nu'\nu''+V''\nu'^2\right]}
 { 9V\left( 1-4\alpha\kappa^4V\right)^2-4\kappa^4VV'^2\nu'^2}.
\end{eqnarray}

In the continuation, we consider the power-law
potential~\eqref{V1} and the Chern-Simons trigonometric coupling
function~\eqref{nu1}. In this case, using the parameter
$\epsilon_1$ of relation~\eqref{NewSRe1,2}, we arrive at
\begin{equation}\label{NewPhiEnd}
\phi_{\rm end}^2\approx \frac{n^2}{2\kappa^2}.
\end{equation}
Also, utilizing relation~\eqref{Neq}, while using
Eqs.~\eqref{NewField11} and~\eqref{NewPhiDot}, we have
\begin{equation}
N(\phi)\approx \frac{\kappa^2}{n}\int^{\phi}_{\phi_{\rm
end}}\dfrac{\phi\, {\rm d}\phi}{1-4\alpha\kappa^4v\phi^n}.
\end{equation}
To find the solution, we can use $\int x\,{\rm
d}x/(1-Ax^n)=[x^2/2]\, {}_2F_1(1,2/n;(n+2)/n;Ax^n)$, where
${}_2F_1(a,b;c;z)=\sum^\infty_{m=0}[(a)_m(b)_m/(c)_m]\, z^m/n!$\,,
with $(q)_m= q(q+1)\cdots (q+m-1)$ for $m>0$ and $(q)_m=1$ for
$m=0$, is the hypergeometric function. At this stage, since the
solution for the unknown $n$ is very complicated, we proceed by
considering $n=2$, which is roughly comparable to the results
obtained in the corresponding linear form of $f(R,T)$ gravity.
Under these considerations, we have
\begin{equation}\label{NewFolding}
N(\phi)\approx\frac{\kappa^2}{2}\left[-\frac{1}{8\alpha\kappa^4
v}\ln\left(1-4\alpha\kappa^4
v\phi^2\right)\right]\bigg|^\phi_{\phi_{\rm end}},
\end{equation}
where in this case $\phi_{\rm end}=2/\kappa^2$. Accordingly, we
obtain
\begin{equation}\label{NewPhiNonL}
\phi^2\approx \frac{D-1+8\alpha\kappa^2 v}{4\alpha\kappa^4vD},
\end{equation}
wherein $\ln D\equiv 16\alpha\kappa^2 v N$.

\begin{figure*}[t!]
\centering \subfigure[]{
\includegraphics[width=0.4\textwidth]{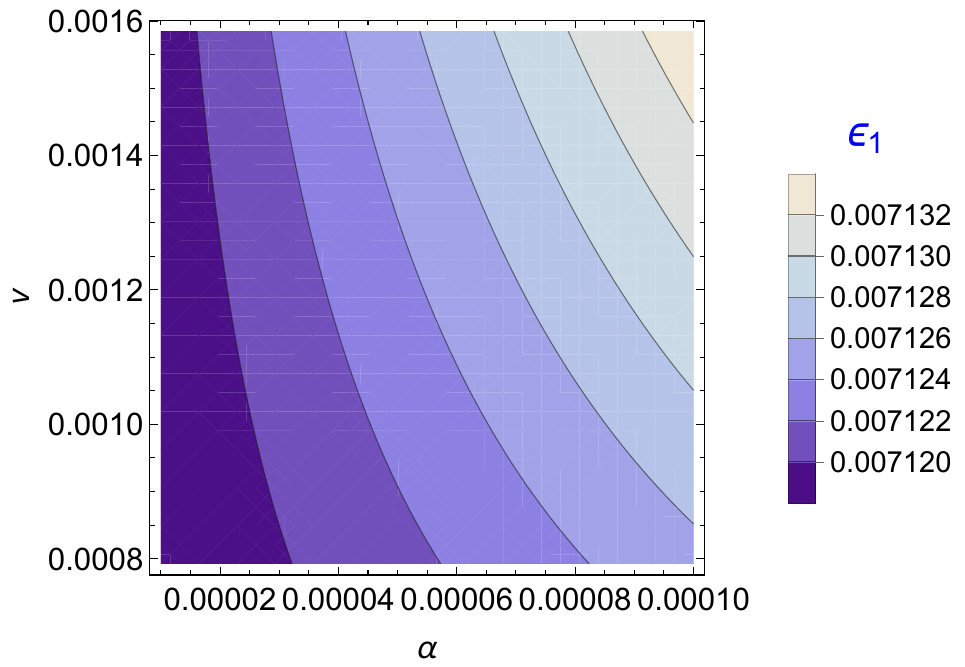}
\label{a1} } \subfigure[] {
\includegraphics[width=0.4\textwidth]{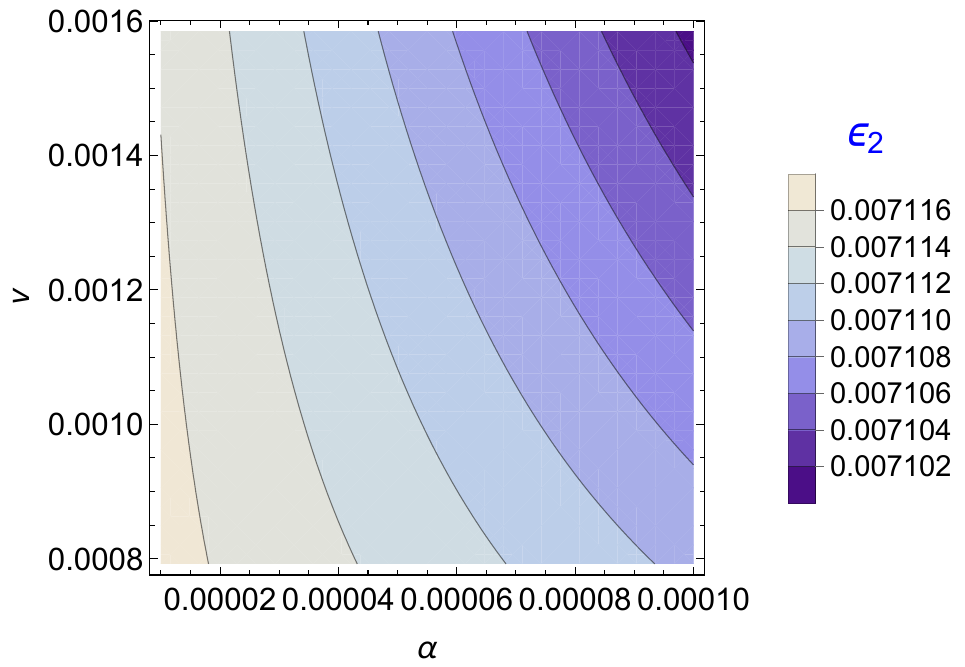}
\label{b1} } \subfigure[] {
\includegraphics[width=0.4\textwidth]{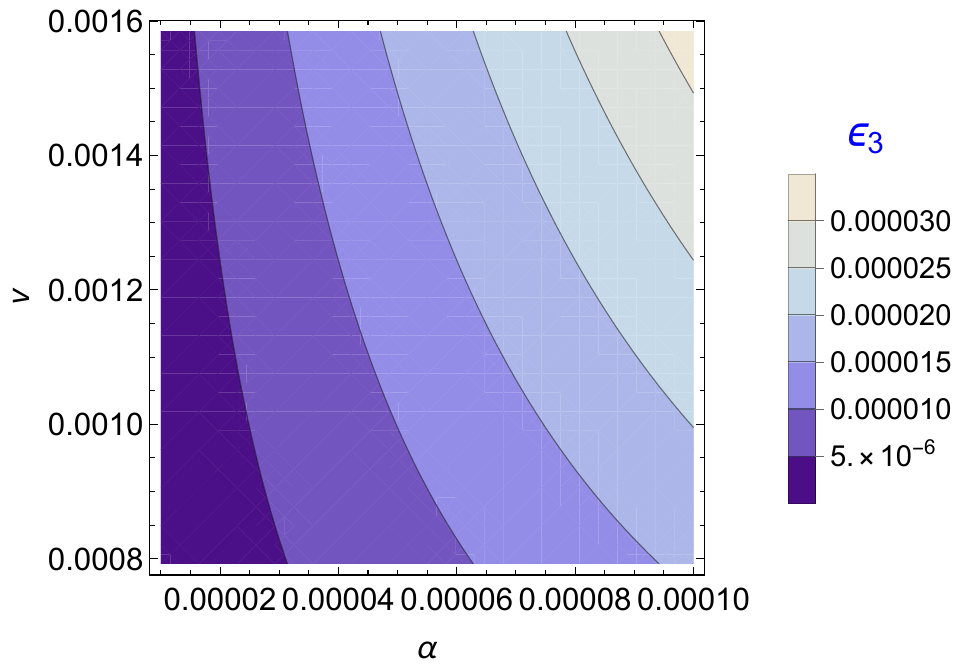}
\label{c1} } \subfigure[] {
\includegraphics[width=0.4\textwidth]{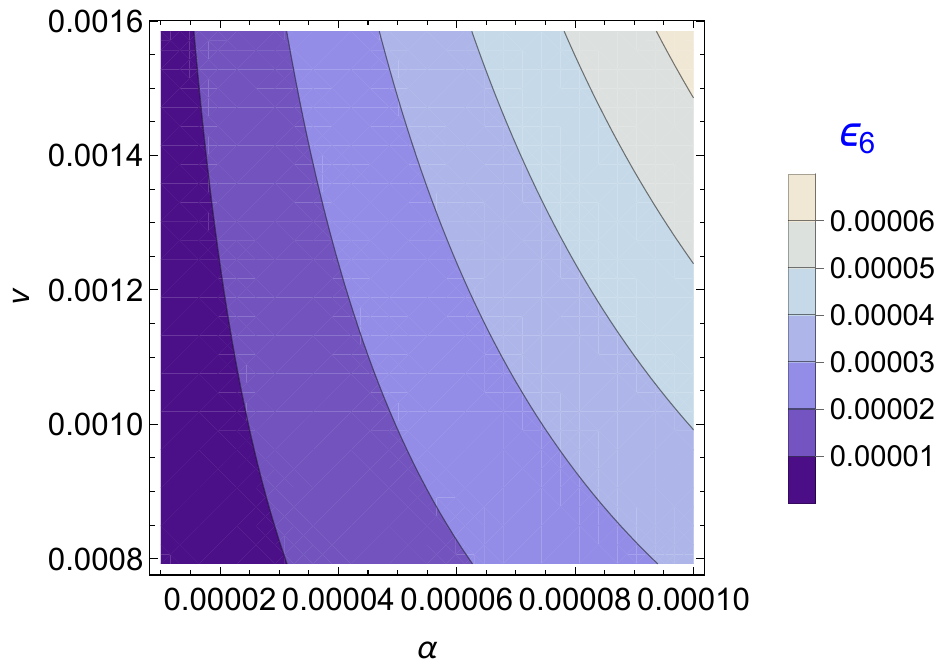}
\label{d1} }
 \caption{\label{nonlinsr} [color online] The slow-roll parameters have been plotted in terms
of the parameters $\alpha$ and $v$, for $N=70$ with the values
$\kappa^2=8 \pi$, $\Lambda=10^{-4}$ and $\phi_1=1/\kappa^2$ in the
range $\alpha=[10^{-5}, 10^{-4}]$ and $v=[0.5/(8\pi)^2,
1/(8\pi)^2]$, where (a) shows the range of parameters
$\epsilon_1$, (b) $\epsilon_2$, (c) $\epsilon_3$ and (d)
$\epsilon_6$.}
\end{figure*}

Thereafter, by inserting relations $V(\phi)=v\phi^2$, $\nu(\phi)$
(relation~\eqref{nu1}) and~\eqref{NewPhiNonL} into the
corresponding relations for the slow-roll parameters
\eqref{NewSRe1,2}, \eqref{NewSRe3,4,5} and \eqref{NewSRe6}, we
obtain those in terms of the model parameters as
\begin{align}
\label{NewEps1} &\epsilon_1\approx \dfrac{8 \alpha \kappa^2 v D}{\big| D-1+8\alpha\kappa^2 v\big|},\\
\label{NewEps2} &\epsilon_2\approx \bigg|\dfrac{4 \alpha \kappa^2 v \left(2-D-8\alpha\kappa^2 v\right)}{D-1+8\alpha\kappa^2 v}\bigg|\\
\label{NewEps3} &\epsilon_3 \approx \epsilon_4 \approx \epsilon_5 \approx 8\alpha \kappa^2 v,\\
\label{NewEps6} &\epsilon_6\approx\nonumber\\
&\bigg|\!\dfrac{16v\left(1\!-\!8\alpha\kappa^2 v\right)\, A}
 {9\alpha\phi_1^3\!\left(1\!-\!8\alpha\kappa^2
 v\right)^2\!\!-\!4v\Lambda^2 \phi_1
 D\left(D\!-\!1\!+\!8\alpha\kappa^2
 v\right)\cos^2\!\left(\!\dfrac{\phi}{\phi_1}\!\right)}\bigg|,
\end{align}
where
\begin{align}
A\equiv &\, 9\alpha^2\kappa^2\phi_1^3\left(1\!-\!8\alpha\kappa^2
  v\right)\!+\!2\alpha\kappa^2v\Lambda^2\phi_1
  \cos^2\!\left(\dfrac{\phi}{\phi_1}\right)\nonumber\\
 &-\!\Lambda^2\sqrt{\alpha v D(D\!-\!1\!+\!8\alpha\kappa^2
v)}\sin\left(\dfrac{\phi}{\phi_1}\right)\cos\left(\dfrac{\phi}{\phi_1}\right).
\end{align}
In Fig.~\ref{nonlinsr}, we have plotted these slow-roll parameters
for $N=70$, $\kappa^2=8\pi$, $\Lambda=10^{-4}$ and
$\phi_1=1/\kappa^2$, as function of $\alpha$ and $v$ parameters in
the range $\alpha=[10^{-5}, 10^{-4}]$ and $v=[0.5/(8\pi)^2,
1/(8\pi)^2]$.  The results show that the slow-roll condition
$|\epsilon_i| \ll 1$ is satisfied, as they should, which also
confirms the internal consistency of the approximation in the
leading-order effects for the dynamics of this non-linear modified
gravity.

\begin{figure*}[t!]
\centering \subfigure[]{
\includegraphics[width=0.4\textwidth]{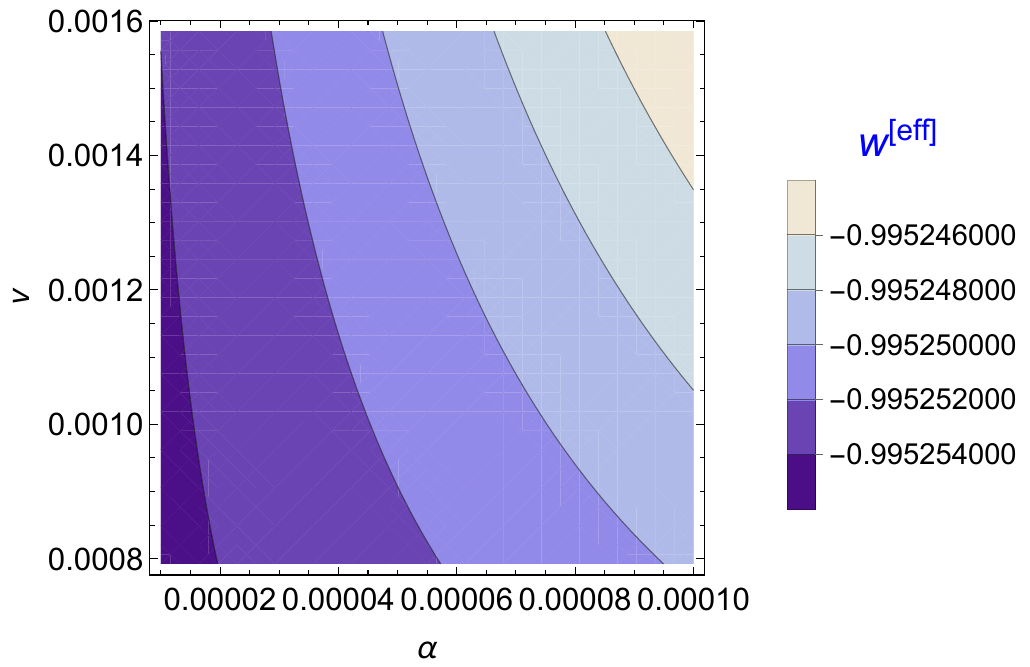}
\label{a1} }
 \caption{\label{nonlinweff} [color online] The effective equation of state parameter during
 inflation has been plotted in terms of the parameters
 $\alpha$ and $v$, for $N=70$ with the value $\kappa^2=8 \pi$ in the
range $\alpha=[10^{-5}, 10^{-4}]$ and $v=[0.5/(8\pi)^2,
1/(8\pi)^2]$.}
\end{figure*}

Also, by using the slow-roll parameter $\epsilon_1$ in
relation~\eqref{NewEps1} into relation \eqref{eos}, we can derive
the effective equation of state parameter for this case in terms
of the parameters $\alpha$ and $v$. Then, in
Fig.~\ref{nonlinweff}, we have plotted it with the same values
mentioned for the other parameters. The figure indicates that the
result is consistent with $w^{\rm [eff]}\approx -1$ during the
inflationary era.

\begin{figure*}[t!]
\centering \subfigure[]{
\includegraphics[width=0.32\textwidth]{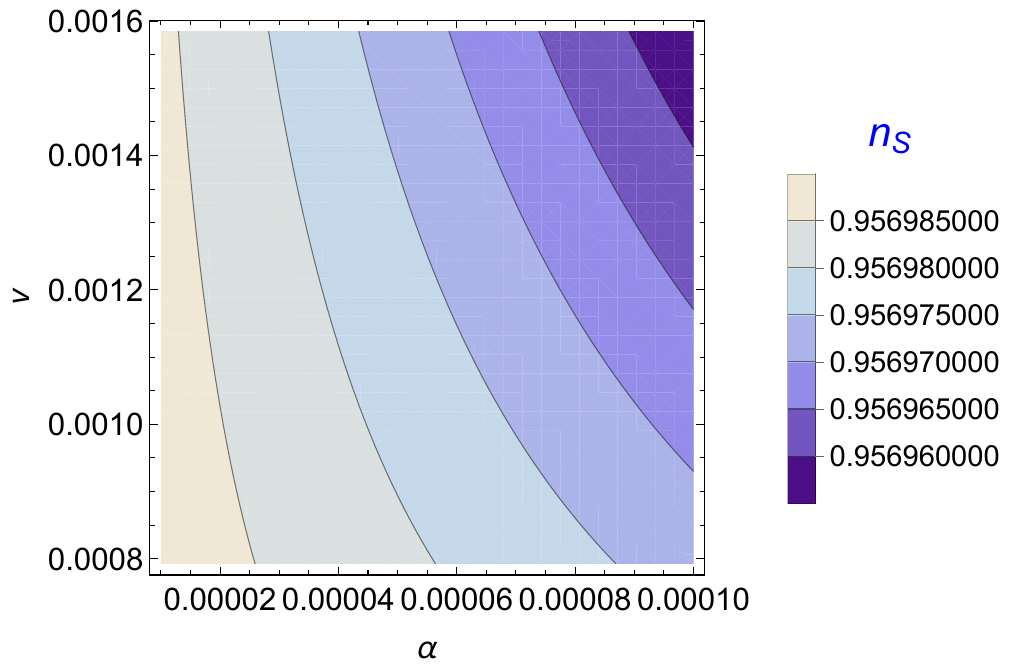}
\label{a1} } \subfigure[] {
\includegraphics[width=0.31\textwidth]{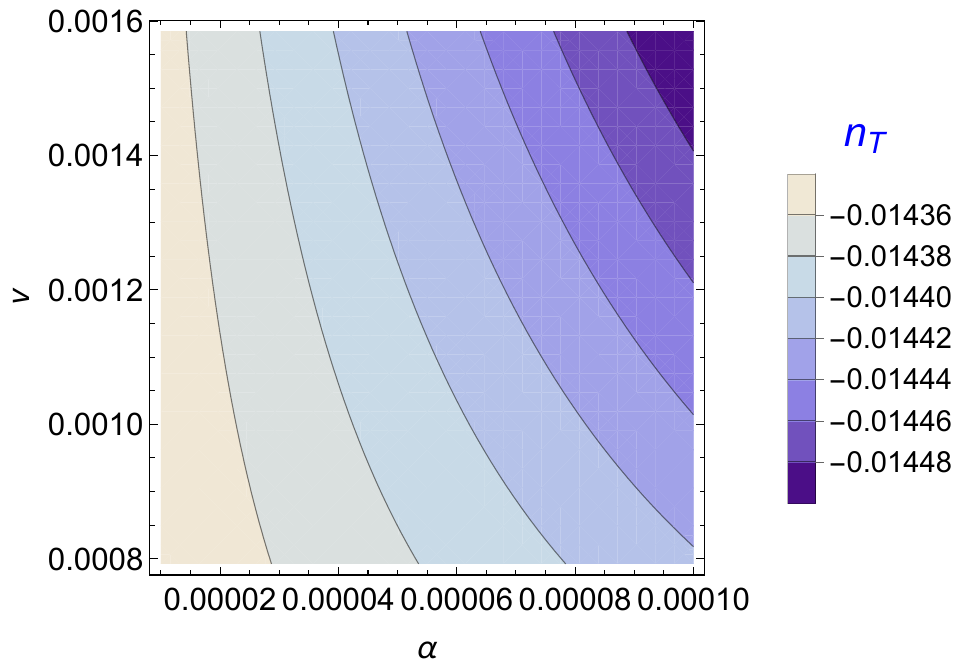}
\label{b1} } \subfigure[] {
\includegraphics[width=0.31\textwidth]{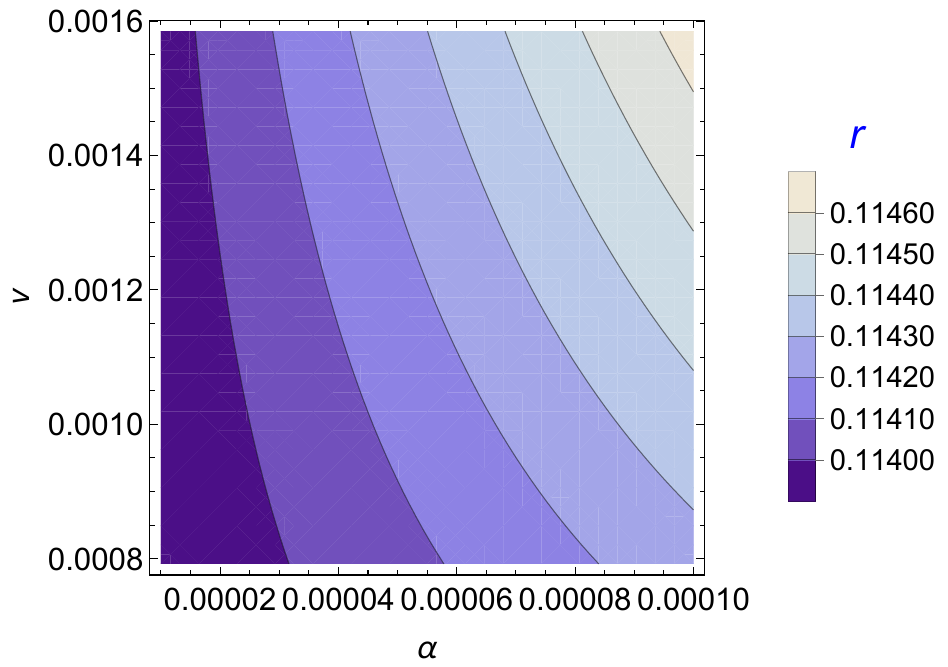}
\label{c1} }
 \caption{\label{nonlinobs} [color online] The inflationary observables have been plotted in terms
of the parameter $\alpha$ and $v$, for $N=70$ with the values
$\kappa^2=8 \pi$, $\Lambda=10^{-4}$ and $\phi_1=1/\kappa^2$ in the
range $\alpha=[10^{-5}, 10^{-4}]$ and $v=[0.5/(8\pi)^2,
1/(8\pi)^2]$, where (a) shows the range of the scalar spectral
index, $n_{\rm S}^{}$, (b) the tensor spectral index, $n_{\rm
T}^{}$, and (c) the tensor-to-scalar ratio, $r$.}
\end{figure*}

Next, by substituting the slow-roll
parameters~\eqref{NewEps1}--\eqref{NewEps6} into
relations~\eqref{nSFirst} and~\eqref{nTFirst}, we obtain the
inflationary observables $n_{\rm S}^{}$ and $n_{\rm T}^{}$ as
functions of the model parameters $\alpha$ and $v$. Meanwhile, in
this case, relation~\eqref{rFirst} for $r$ is obtained as
\begin{equation}\label{New}
r\approx \dfrac{144(\epsilon_1+\epsilon_3)\left(1-4\alpha\kappa^4
V\right)^2}{9\left(1-4\alpha\kappa^4 V\right)^2-4\kappa^4
V'^2\nu'^2}.
\end{equation}
Then, by inserting relations $V(\phi)=v\phi^2$, \eqref{nu1},
\eqref{NewEps1} and \eqref{NewEps3}, while using relation
\eqref{NewPhiNonL}, into relations~\eqref{nSFirst}, we also obtain
$r$ as a function of $\alpha$ and $v$. Accordingly, in
Fig.~\ref{nonlinobs}, we have plotted these three inflationary
observables in terms of the parameters $\alpha$ and $v$ and for
the specified values of the other parameters, all adjusted to
achieve the most favorable outcomes. The plots in
Fig.~\ref{nonlinobs} indicate results that are approximately
compatible with the Planck data~\eqref{planckdata}. However, in
comparison, the obtained results show that the corresponding
linear form of $f(R,T)$ gravity provides a slightly better
prediction, for $r$, than this non-linear form of $f(R,T)$ gravity
in terms of consistency with the Planck data~\eqref{planckdata}.

\section{Conclusions}
We have investigated cosmological inflation in the framework of
the linear form of $f(R,T)$ gravity with a minimal coupling
parameter that quantifies the strength of the matter-curvature
interaction. With this choice of the linear form, three of the
slow-roll parameters (i.e., $\epsilon_3$, $\epsilon_4$ and
$\epsilon_5$) vanish and do~not contribute to the corresponding
inflationary dynamics, which simplifies the calculations.
Furthermore, the matter coupling parameter alters the effective
energy density and effective pressure in the Friedmann equations,
leading to modified evolution of the Hubble parameter and the
inflaton field, and in turn the slow-roll parameters and the
inflationary observables. In general, by incorporating the
matter-curvature coupling, $f(R,T)$ gravity provides a more
detailed description of inflationary dynamics that accounts for
the effect of matter field on spacetime curvature.

We have also incorporated the inflaton scalar field and the
Chern-Simons higher-curvature corrections induced by aspects of
quantum gravity. Then, utilizing the FLRW metric, we have derived
the modified Friedmann equations. Thereafter, by employing the
slow-roll conditions and defining the necessary slow-roll
parameters, we have acquired their required expressions in the
proposed model. In the continuation, we have focused on two
specific cases for the the Chern-Simons coupling function,
trigonometric and exponential, along with two different forms of
the inflaton potential. Such coupling functions have been chosen
based on their ability to introduce non-trivial effects into
inflationary dynamics while remaining analytically manageable. The
trigonometric form has been selected to explore periodic
variations in the coupling, whereas the exponential form has been
adopted to examine the implications of rapidly decaying coupling
functions\rlap.\footnote{It should be noted that a linear form of
coupling function, as a more natural one, would preserve the shift
symmetry. In this regard, in Ref.~\cite{Venikoudis:2021oee},
adopting a linear form for the Chern-Simons coupling leads to a
`trivial' scenario that remains compatible with the observational
data. However, the study of shift symmetry can lead to interesting
phenomena such as the asymmetric production of circularly
polarized gravitational waves that serves as a motivation for
future studies.}\
 Accordingly, we have obtained the
scalar spectral index, tensor spectral index, and tensor-to-scalar
ratio in this modified gravity model.

In the absence of the Chern-Simons correction, the $\epsilon_6$
slow-roll parameter is zero. However, including the Chern-Simons
correction introduces a parity-violating term in the gravitational
action and leads to a non-zero value for the $\epsilon_6$
slow-roll parameter, which alters the tensor perturbation dynamics
during inflation. This approach modifies the gravitational wave
power spectrum, affecting the tensor spectral index and changing
the tensor-to-scalar ratio. The key difference arises from the
additional higher-curvature terms introduced by the Chern-Simons
correction, which provide new insights into the interplay between
parity-violating effects and inflationary dynamics that can
potentially revive some inflationary scenarios. However, we remind
that there is non-participation of the Chern-Simons correction in
the field equations and the scalar perturbations. Indeed, the
phenomenological treatment of the Chern-Simons correction requires
a deeper theoretical grounding from the point of view of an
ultraviolet-complete quantum gravity framework.

In the first case, i.e. the trigonometric coupling function and
the power-law potential, by constraining the free parameters of
the model, i.e. the potential power and the e-folding number
within certain ranges, while fixing the values of the rest of
constants, we have first plotted the slow-roll parameters to
ensure that their values used are suitable to satisfy the
slow-roll condition. Then, we have plotted the inflationary
observables with the same values. As a result, we have found that
this model yields predictions well-aligned with the observed
scalar spectral index and the upper bounds on the tensor-to-scalar
ratio from the Planck $2018$ data. Also, the plot of the results
in the $(r, n_{\rm S}^{})$ space and for the effective equation of
state parameter highlights that the configuration with the
lower-power for the inflaton potential, i.e. $n = 1.5$, and the
acceptable range of the e-folding number, i.e. $60<N\leq 70$,
provides the closest agreement with observational data.

In the second case, i.e. the exponential coupling function and a
type of hilltop potential, by constraining the free parameters of
this model, i.e. the parameter of the linear form of $f(R,T)$
gravity (the $\beta$ parameter) and the parameter of the potential
function (the $\gamma$ parameter) within certain ranges, and
fixing the e-folding number, once again we have first plotted the
obtained slow-roll parameters to ensure that the values used in
plotting them are suitable to satisfy the slow-roll condition.
Then, we have plotted the inflationary observables with the same
values. As a result, the obtained findings indicate that this
model can also lead to well-aligned results with the Planck $2018$
data and the combined Planck, BK$15$ and BAO data. Also, the plot
of the results in the $(r, n_{\rm S}^{})$ space and for the
effective equation of state parameter indicates that the
configuration employing distinct values of the $\beta$ and
$\gamma$ parameters with the acceptable value of the e-folding
number aligns well with those observational constraints. Moreover,
the results obtained show that incorporating the Chern-Simons
correction with its exponential coupling function enables the
$f(R, T)$ gravity to impose more stringent constraint on the upper
bound of the tensor-to-scalar ratio than the model that utilizes
the Chern-Simons trigonometric coupling function.

To explore the motivation for considering the presented model, as
well as for comparative analysis and better comparison, we have
examined the model without the Chern-Simons correction, the model
without the linear form of $f(R,T)$ gravity (i.e., with the trace
of the energy-momentum tensor removed), and the non-linear form of
$R+\alpha\kappa^4 R\, T$ gravity plus the inflaton scalar field in
the presence of the Chern-Simons correction. We have investigated
the first two cases for both inflaton potentials, each with the
corresponding Chern-Simons coupling function, and the latter one
for the power-law potential with the power of two and the
Chern-Simons trigonometric coupling function. In almost all five
cases, the plots drawn and the results obtained indicate only
slight deviations from the predictions of the corresponding ones
from the presented model. Nevertheless, we have examined the
differences between some closely related parameters of each case
with the related one in the presented model, in more detail,
through calculating their ratios. Almost all the ratios obtained
show that their difference with $1$ is negligible. Therefore, and
as a general conclusion, the findings obtained indicate that the
inclusion of the Chern-Simons correction approximately refines the
values of the tensor spectral index and tensor-to-scalar ratio in
the context of the linear form of $f(R,T)$ gravity.

Finally, we emphasize that the applicability of the usual
slow-roll approximations, as adopted within the presented $f(R,T)$
gravity with Chern-Simons correction, has been verified. Actually,
from a numerical perspective, the sufficiently small values
obtained for the slow-roll parameters confirm the internal
consistency of the approximation in the leading-order effects for
the dynamics of this modified gravity model. Furthermore, this
model is efficient because the Chern-Simons correction does~not
contribute to the background Friedmann equations. Nevertheless,
although the current model yields a consistent slow-roll
inflationary scenario, extending the present analysis beyond the
leading-order slow-roll treatment to more general modified gravity
frameworks could potentially provide further insights and be an
interesting direction for future investigations.

\section*{ACKNOWLEDGMENTS}
The authors thank the Deputy for Research and Technology of Shahid
Beheshti University. Also, SF acknowledges financial support from
the Generalitat Valenciana and the European Social Fund through
the APOSTD 2025 postdoctoral fellowship (CIAPOS/2024/461), the
Spanish Agencia Estatal de Investigaci\'{o}n (grant
PID2024-159689NB-C21, funded by MICIU/AEI/10.13039/501100011033
and FEDER/EU), the Generalitat Valenciana (Prometeo grant
CIPROM/2022/49), and the European Horizon Europe Staff Exchanges
programme (Grant No. NewFunFiCO-101086251).


%
\end{document}